\documentclass[12pt, draft, onecolumn]{IEEEtran}
\def\figs{7 cm}
\def\figswide{9 cm}
\usepackage{epsf}
\usepackage{ifthen}
\usepackage{times}

\newcommand{\thd}[1]
{\ifthenelse {\equal{#1}{1}}
	{{#1}$^{\mathrm{st}}$} 
	{{\ifthenelse{\equal{#1}{2}}{{#1}$^{\mathrm{nd}}$}{{#1}$^{\mathrm{th}}$}}}}

\newcommand{\CAWGN}{C_\mathrm{AWGN}}
\newcommand{\SNR}{\mathrm{SNR}}

\newtheorem{Theorem}{Theorem}
\newtheorem{Theorem_converse}[Theorem]{Converse to Theorem}
\newtheorem{Theorem_nonum}[Theorem]{Theorem}
\newtheorem{Part}{Part}
\newtheorem{Converse}{Converse}

\newcounter{tempCounter}

\usepackage{amsmath}

\newtheorem{Definition}{Definition}
\newtheorem {Lemma} [Theorem]    {Lemma}
\newtheorem {Proposition}[Theorem]    {Proposition}

\newcommand{\bz}{{\bf Z}}
\newcommand{\bp}{{\bf P}}
\newcommand{\pnl}{{\bf P}_{L_m,L}}
\newcommand{\note}{\noindent {\bf Note. }}

\begin{document}

\title{Channel Uncertainty in Ultra Wideband Communication Systems}
\author{Dana Porrat, David N. C. Tse and Serban Nacu%
\thanks{
Presented at the 41$^{\mathrm{st}}$ Allerton Conference on Communication, Control and Computing, October 2003.}
\thanks{
Dana Porrat is with the Hebrew University, dporrat@cs.huji.ac.il.
David Tse is with the University of California at Berkeley, dtse@eecs.berkeley.edu.
Serban Nacu was with the University of Clifornia at Berkeley at the time of writing, serban@stat.berkeley.edu.}
\thanks{
This work was funded in part by the Army Research Office (ARO) under grant \#DAAD19-01-1-0477, via the University of Southern California.
}%
}
\maketitle

\begin{abstract}
Wide band systems operating over multipath channels may spread their power over bandwidth if they use duty cycle.
Channel uncertainty limits the achievable data rates of power constrained wide band systems;
Duty cycle transmission reduces the channel uncertainty because the receiver has to estimate the channel only when transmission takes place.
The optimal choice of the fraction of time used for transmission depends on the spectral efficiency of the signal modulation.
The general principle is demonstrated by comparing the channel conditions that allow different modulations to achieve the capacity in the limit.
Direct sequence spread spectrum and pulse position modulation systems with duty cycle achieve the channel capacity, if the increase of the number of channel paths with the bandwidth is not too rapid.
The higher spectral efficiency of the spread spectrum modulation lets it achieve the channel capacity in the limit, in environments where pulse position modulation with non-vanishing symbol time cannot be used because of the large number of channel paths.

\end{abstract}


\section{Introduction}

This work discusses the achievable data rates of systems with very wide bandwidths.
Considering communication with an average power constraint, the capacity of the multipath channel in the limit of infinite bandwidth is identical to the capacity of the additive white Gaussian noise (AWGN) channel \mbox{$C_\mathrm{AWGN}=P/N_0 \log e$}, where $P$ is the average received power (due to the transmitted signal) and $N_0$ is the received noise spectral density.
{\em Kennedy}~\cite{kennedy_book} and {\em Gallager}~\cite{gallager_book_sec8.6} proved this for fading channels using FSK signals with duty cycle transmission; {\em Telatar \& Tse}~\cite{telatar_2000} extended the proof for multipath channels with any number of paths.
The AWGN capacity is achievable on multipath channels also by dividing the spectrum into many narrow bands and transmitting bursty signals separately on each band.

When using spreading modulations, {\em M\'{e}dard \& Gallager}~\cite{medard_2002} show that direct sequence spread spectrum signals, when transmitted continuously (no duty cycle) over fading channels (that have a very large number of channel paths), approach zero data rate in the limit of infinite bandwidth.
A similar result was shown by {\em Subramanian \& Hajek}~\cite{subramanian_2002}.
{\em Telatar \& Tse}~\cite{telatar_2000} show that over multipath channels, the data rate in the limit of infinite bandwidth is inversely proportional to the number of channel paths.

This work is motivated by a recent surge in interest in ultra wide band systems, where spreading signals are often desired.
It shows that under suitable conditions, spreading signals can achieve AWGN capacity on multipath channels in the limit of infinite bandwidth, if they are used with duty cycle.
In other words, peakiness in time is sufficient to achieve AWGN capacity, and the transmitted signal does not have to be peaky in frequency as well.
We analyze direct sequence spread spectrum (DSSS) and pulse position modulation (PPM) signals, and show that when the scaling of the number of channel paths is not too rapid, these signals achieve the capacity in the limit as the bandwidth grows large.
DSSS signals with duty cycle were used by~\cite{zheng_OnChannel_2003}, that showed non-vanishing capacity in the limit.

Our results can be seen as a middle ground between two previous results: 1.\ FSK with duty cycle achieves AWGN capacity for any number of channel paths and 2.\ direct sequence spread spectrum signals with continuous transmission (no duty cycle) have zero throughput in the limit, if the number of channel paths increases with the bandwidth.

The effect of duty cycle can be understood in terms of the channel uncertainty a communication system faces.
The data rate is penalized when the receiver has to estimate the channel, so infrequent usage of the channel leads to a small channel uncertainty and a small penalty.
The spectral efficiency of the modulation scheme plays an important role in determining the channel uncertainty a system handles.
A system with a low spectral efficiency can pack a small number of bits into each transmission period, and in order to maintain a high data rate it must transmit often.
Thus, low spectral efficiency forces the communication system to estimate the channel often, and suffer from a large penalty on its data rate.

A useful intuition is gained from examining the ratio
\[
\mathrm{SNR}_\mathrm{est}=\frac{P}{N_0}\frac{T_c}{\theta L}
\]
where $L$ is the number of independent channel components, $T_c$ is the coherence time and $\theta$ is the duty cycle parameter or the fraction of time used for transmission.
The channel uncertainty (per channel realization) depends linearly in our model on $L$, and normalizing it to unit time means dividing by the coherence time $T_c$ and multiplying by the fraction of time used for communication.
We thus see that $\SNR_\mathrm{est}$ compares the channel uncertainty per unit time $\frac{\theta L}{T_c}$ to the data rate in the limit of infinite bandwidth that is proportional to $\frac{P}{N_0}$.
The ratio $\mathrm{SNR}_\mathrm{est}$ can also be interpreted as the effective SNR per path for channel estimation.
A communication system can achieve the channel capacity in the limit of infinite bandwidth if channel estimation in the limit is perfect, and this requires 
\[
\mathrm{SNR}_\mathrm{est}\rightarrow\infty
\]
In systems with bounded average received power, the duty cycle parameter must diminish in order to balance the increase in the number of channel components $L$, and let the overall channel uncertainty diminish.
This work examines the conditions that allow this situation in the limit, for two different modulations with different spectral efficiencies.
The combination of the requirements for diminishing channel uncertainty ($\theta L\rightarrow0$) and sufficient transmitted bit rate determines the conditions that allow the data rate to converge to the channel capacity in the limit, for systems where the receiver knows the path delays but not their gains.
The requirement for sufficient data rate relates the bandwidth $W$ to the duty cycle parameter, and depends on the spectral efficiency of the modulation scheme.
Spectrally efficient modulation schemes permit infrequent transmission (or small $\theta$), thus reducing the channel uncertainty per unit time.
In contrast, low spectral efficiency forces frequent transmission, and the duty cycle parameter must stay high.

The difference between the wideband capacities of DSSS and PPM schemes comes about precisely because of their different spectral efficiencies.
PPM is an orthogonal modulation, so the number of bits it can transmit per unit time increases only logarithmically with the bandwidth, where the number of bits a DSSS transmitter can send per unit time increases linearly.
Thus, DSSS can tolerate a  larger amount of channel uncertainty than PPM.
Note that, in contrast, both PPM and DSSS achieve the channel capacity in the limit of infinite bandwidth over the AWGN channel as well as over the multipath fading channel where the channel is known to the receiver.

\begin{figure}
\begin{center}
\leavevmode
\epsfysize=\figs \epsfbox{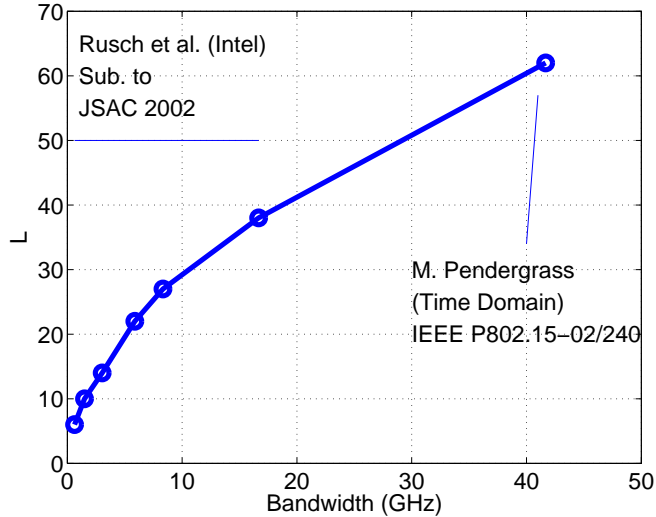}
\caption{
Number of significant channel paths vs.\ bandwidth.
The paths accounting for 60--90 percent of the energy were counted in two separate measurement campaigns.
The Intel data were taken from~\cite{rusch_2002} and the Time Domain data from~\cite{pendergrass_2002}
}
\label{fig:LvsW}
\end{center}
\end{figure}

Our main results are as follows.
In the limit of infinite bandwidth, DSSS systems where the receiver knows the path delays achieve AWGN capacity if the number of channel path is sub--linear in the bandwidth, formally if $\frac{L}{W}\rightarrow0$ where $L$ is the number of independently fading channel paths and $W$ is the bandwidth, and the system uses an appropriate duty cycle.
PPM systems too can achieve AWGN capacity in the limit of infinite bandwidth, but this is possible for smaller number of channel paths.
A PPM system with a receiver that knows the path delays achieves AWGN capacity if $\frac{L}{\log W}\rightarrow0$.
PPM systems with lower bounded symbol time have zero throughput if $\frac{L}{\log W}\rightarrow\infty$.
In systems where the receiver does not know the path gains or delays, we show that DSSS systems can achieve AWGN capacity if $\frac{L}{W/\log W}\rightarrow 0$ as the bandwidth increases.
Measurements of the number of channel paths vs.\ bandwidth in Figure~\ref{fig:LvsW} show an increase of the number of channel paths that appears to be sub--linear.

It is interesting to contrast our results with those of Verd\'{u}~\cite{verdu_2002}, where the crucial role of peakiness in time is pointed out and a connection is also made between the spectral efficiency of a modulation scheme and its data rate over a large bandwidth.
The theory there analyzes the capacity of channels with {\em fixed} channel uncertainty as the average received power per degree of freedom goes to zero (or, equivalently, the bandwidth goes to infinity).
By suitable scaling of the duty cycle, the infinite bandwidth AWGN capacity is always approached in the limit by the modulation schemes considered there, and the issue is the {\em rate of convergence} to that limit.
This rate of convergence depends on the spectral efficiency of the modulation scheme.
In contrast, the environment considered here is a harsher one, in that the channel uncertainty (or the number of channel parameters) {\em increases} with the bandwidth.
Now the issue is {\em whether} a modulation scheme approaches the AWGN capacity at all.
The framework in~\cite{verdu_2002} is suitable for systems that break the wideband channel into many parallel narrow-band channels; one example is the orthogonal frequency division multiplexing (OFDM) modulation.
In this context, one can focus on a single narrowband channel, in which the channel uncertainty is fixed.
The channel uncertainty faced by an OFDM system depends on the number of such parallel channels, it does not depend on the number of channel paths $L$.

Analysis of the zero SNR limit over a narrowband channel is also offered by {\em Zheng et al}.
In~\cite{zheng_Channel_2004,zheng_OnThe_2004} they analyze the channel uncertainty penalty of narrowband systems in the limit of diminishing SNR.
The concept of SNR per degree of freedom of the channel is central in these analyses, that are based on the connection between the coherence time and the SNR per degree of freedom.
In the narrowband setup, each degree of freedom of the channel is a single (scalar) channel realization.
The longer the coherence time, the more energy is transmitted per block, thus the SNR per degree of freedom increases and estimation of the (scalar) channel improves accordingly.

The outline of the paper is as follows.
After presenting the channel model and the signals in Section~\ref{sec:setup}, a summary of the results is presented in Section~\ref{sec:summary} and discussion is brought in Section~\ref{sec:discussion}.
Section~\ref{sec:penalty} presents a bound on the channel uncertainty penalty related to path delays.
Sections~\ref{sec:DSSS} and~\ref{sec:ppm} then present bounds on the data rates of DSSS and PPM system where the receiver knows the channel path delays.

\section{Channel Model and Signal Modulations}
\label{sec:setup}

The natural model for an ultra wide band channel is real, because there is no obvious `carrier frequency' that defines the phase of complex quantities.
The channel is composed of $\tilde{L}$ paths:
\[
Y(t)=\sum_{l=1}^{\tilde{L}}A_l(t)X(t-d_l(t))+Z(t)
\]
The received signal $Y(t)$ is match filtered and sampled at a rate $1/W$, where $W$ is the bandwidth of the transmitted signal, yielding a discrete representation of the received signal.
We assume a block fading model: the channel remains constant over coherence periods that last $T_c$, and changes independently between coherence periods.
The paths' delays $\left\{d_l\right\}_{l=1}^L$ are in the range $[0,T_d)$, where $T_d$ is the delay spread.
The channel is assumed under-spread, so the delay spread $T_d$ is much smaller than the coherence period, and signal spillover from one coherence period to the next is negligible.
Considering a double sided noise density $\frac{N_0}{2}$ the discretized and normalized signal is given by
\begin{equation}
Y_i=\sqrt{\frac{\mathcal{E}}{K_c}}\sum_{m=1}^{\tilde{L}} A_m X_{i-\tau_m}+Z_i\ \ \ \ \ i=0,\dots,\lfloor T_cW\rfloor-1 \label{eq:chmodel}
\end{equation}
with $\mathcal{E}=\frac{2PT_c}{N_0\theta}=2L\mathrm{SNR}_\mathrm{est}$ and $K_c=\lfloor T_cW\rfloor$.
The noise $\{Z_i\}$ is real and Gaussian, and the normalization requires that the path gains $\left\{A_m\right\}$ and the transmitted signal $\left\{X_i\right\}$ are scaled so that $E\left(\sum A_m X_{i-m}\right)^2=1$.
We assume that $\left\{A_m\right\}$ are independent and zero mean, thus the scaling requirement is $\sum E\left[ A_m^2\right]=1$ and $E\left[X_i^2\right]=1$.
This normalization ensures that $E[Z_i^2]=1$.
$P$ is the average received power and $W$ is the bandwidth.
In practical systems, where the channel response is dispersive, the bandwidth of the received signal is larger than the bandwidth of the transmitted signal. 
We do not consider this difference in the channel model.

In order to avoid complications at the edge of the coherence interval, we approximate the channel using a cyclic difference over $K_c$ instead of a simple difference: $(n)\equiv n\ \mathrm{mod}\ K_c$.
The difference is negligible as the delay spread is much smaller than the coherence time.
\begin{equation}
Y_i=\sqrt{\frac{\mathcal{E}}{K_c}}\sum_{m=1}^{\tilde{L}} A_m X_{\left(i-\tau_m\right)}+Z_i\ \ \ \ \ i=0,\dots,\lfloor T_cW\rfloor-1 \label{eq:chmodel_circ}
\end{equation}
Note that when $X$ is a PPM signal described in Section~\ref{sec:PPM_signal} (that includes a $T_d$ guard time ensuring a silence period between symbols) the circular formulation~(\ref{eq:chmodel_circ}) is identical to the original model~(\ref{eq:chmodel}).

The path delays $\left\{\tau_m\right\}$ are grouped into resolvable paths, separated by the system time resolution $\frac{1}{W}$.
The resolvable paths are defined by summing over the paths with similar delays:
\[
G_l=\sum_{m:\frac{l}{W}\leq \tau_m< \frac{l+1}{W}} A_m \ \ \ \ \ 0\leq l < \lfloor T_dW \rfloor
\]
The number of resolvable paths is $L$ and their delays $\left\{D_l\right\}_{l=1}^L$ are integers between $0$ and $\lfloor WT_d\rfloor-1$.
The $L$ delays are uniformly distributed over the $\binom{\lfloor WT_d\rfloor}{L}$ possibilities of combinations of $L$ values out of $\binom{\lfloor WT_d\rfloor}{L}$ positions.
\[
Y_i=\sqrt{\frac{\mathcal{E}}{K_c}}
\sum_{l=1}^{L} G_l X_{i-D_l}+Z_i\ \ \ \ \ i=0,\dots,\lfloor T_cW\rfloor-1
\]
The channel gains are real, we assume that they are IID and independent of the delays.
Our normalization requires that the variance of the gains equals $\frac{1}{L}$.

The systems we consider do not use channel information at the transmitter, and the receiver knows the deterministic features of the channel, namely the coherence time $T_c$, the delay spread $T_d$ and the number of paths $L$.
The receiver sometimes has additional knowledge, about the random features of the channel, in particular the path delays $D$.
The assumptions on receiver knowledge of the channel are explicitly stated throughout the paper.

The assumptions we make of IID paths that compose the channel ensure that the channel uncertainty is linearly dependent on the number of paths $L$.
Our results essentially compare the communication data rate to the information (number of bits) needed to describe the channel.
Complex channel models, with non-uniform paths, require more delicate analysis to evaluate the channel uncertainty penalty.

\subsection{Direct Sequence Spread Spectrum Signals}

\begin{figure}
\begin{center}
\leavevmode
\epsfxsize=\figswide \epsfbox{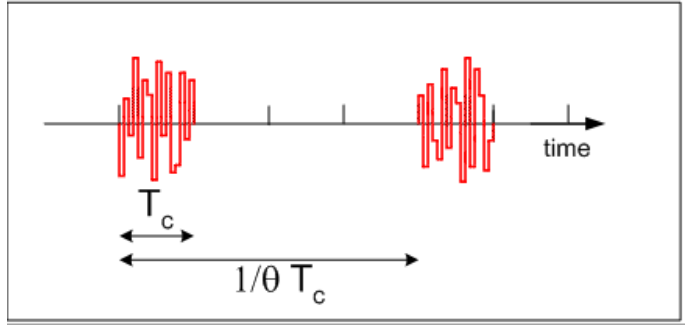}
\caption{
Direct sequence spread spectrum with duty cycle over coherence periods. 
The duty cycle parameter $\theta$ equals the fraction of time used for transmission.
The receiver is aware of the active periods of transmission.
}
\label{fig:DSSS_timing}
\end{center}
\end{figure}

Each transmitted symbol contains a random series of IID $K_c$ Gaussian values $\left\{r_i\right\}_{i=0}^{K_c-1}$ with zero mean, and an energy constraint is satisfied:
\[
E\left[X_j^2\right]=\theta E\left[ \frac{1}{K_c}\sum_{i=0}^{K_c-1} r_i^2\right] = 1
\]
where $0<\theta\leq1$ is the duty cycle parameter or the fraction of time used for transmission.
The symbol $r$ is used for the transmitted signal during active transmission periods, while $X$ represents all of the transmitted signal, that includes silent periods due to duty cycle.
The duty cycle, with parameter $\theta$, is used over coherence times:
of each period $\frac{T_c}{\theta}$, one period of $T_c$ is used for transmission and in the rest of the time the transmitter is silent (Figure~\ref{fig:DSSS_timing}).
We define the sample autocorrelation of the signal
\[
C(m,n) \equiv \frac{\theta}{K_c} \sum_{i=0}^{K_c-1} r_{i-m}r_{i-n} \ \ \ \ \ \ \ \ \ \forall m,n 
\]
Edge conditions are settled by assuming that the each symbol follows a similar and independent one.
Under the assumption of IID chips we have
\begin{equation}
\sum_{m=1}^L \sum_{n=1}^L E_x \left| C(m,n)-\delta_{mn}\right| \leq  \frac{2L}{\sqrt{\pi K_c}} +\frac{L^2-L}{K_c} \label{eq:dsss_iid_cond} 
\end{equation}
see the proof in the appendix.

The upper bound on DSSS capacity (Section~\ref{sec:DSSS_ub}) is also valid for another type of signals, where $x$ is composed of pseudo-random sequences of $K_c$ values.
The empirical autocorrelation of the input is bounded and the signal has a delta--like autocorrelation:
\begin{equation}
\left| C(m,n)-\delta(n,m)\right|\leq\frac{d}{K_c} \label{eq:assumption2}
\end{equation}
where $d$ does not depend on the bandwidth.

\subsection{PPM Signals}
\label{sec:PPM_signal}
The signals defined in this section are used to calculate lower bounds on the data rates of PPM systems (Section~\ref{sec:PPM_dknown}).
The upper bound on PPM performance holds for a wider family of signals, defined in Section~\ref{sec:PPM_ub}.

\begin{figure}
\begin{center}
\leavevmode
\epsfxsize=\figswide \epsfbox{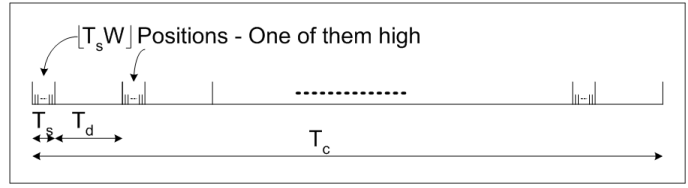}
\caption{
PPM symbol timing.
}
\label{fig:PPM_timing}
\end{center}
\end{figure}

Signaling is done over periods $T_s$ long, with $\lfloor T_sW\rfloor $  positions in each symbol, as illustrated in Figure~\ref{fig:PPM_timing}.
A guard time of $T_d$ is taken between the symbols, so the symbol period is $T_s+T_d$.
The symbol time $T_s$ is typically in the order of the delay spread or smaller.
It does not play a significant role in the results.
The transmitted signal over a single coherence period is 
\begin{eqnarray*}
r_i & = & \left\{ \begin{array}{ll} \sqrt{W\left(T_s+T_d\right)} & \mathrm{one\ position\ of\ each\ group\ of\ }\lfloor T_sW\rfloor \\ 
& \mathrm{with\ } \scriptstyle{n\lfloor (T_s+T_d) W\rfloor \leq i \leq n\lfloor (T_s+T_d) W\rfloor+\lfloor T_s W\rfloor-1} \\
& n=0,1,\dots, \left\lfloor \frac{T_c}{T_s+T_d}\right\rfloor-1\\
\\
0 & \mathrm{other\ positions}
\end{array} \right. \\
&& i=0,1,\dots,\left\lfloor T_cW \right\rfloor-1 
\end{eqnarray*}
The number of symbols transmitted over a single coherence period is \mbox{$N=\frac{T_c}{T_s+T_d}$}.
We assume $N$ is a whole number, this assumption does not alter the results we prove here.
The duty cycle parameter $0<\theta\leq 1$ is used over coherence periods:
of each period $\frac{T_c}{\theta}$, one period of $T_c$ is used for transmission and in the rest of the time the transmitter is silent.

\section{Summary of the Results}
\label{sec:summary}

A direct sequence spread spectrum system with receiver knowledge of the path delays, can tolerate a sub--linear increase of the number of paths with the bandwidth, and achieve AWGN capacity (Section~\ref{sec:DSSS_lb_dknown}).
Conversely, if the number of paths increases linearly with the bandwidth, the data rate is penalized (Sections~\ref{sec:DSSS_ub}).
\begin{Theorem}
\label{th:dsss}
\begin{Part}
\label{prt:dsss_dknown}
DSSS systems with duty cycle where the receiver knows the path delays (but not necessarily their gains) achieve \mbox{$C_\mathrm{DSSS}\rightarrow C_\mathrm{AWGN}$} as \mbox{$W\rightarrow\infty$} if \mbox{$\frac{L}{W}\rightarrow 0$}.
\end{Part}
\begin{Part}
\label{prt:dsss_dunknown}
DSSS systems with duty cycle achieve \mbox{$C_\mathrm{DSSS}\rightarrow C_\mathrm{AWGN}$} as \mbox{$W\rightarrow\infty$} if \mbox{$\frac{L\log{W}}{W}\rightarrow 0$}.
No knowledge of the channel is required.
\end{Part}
\begin{Converse} DSSS systems with duty cycle where the path gains are unknown to the receiver and uniformly bounded by \mbox{$|G_l|\leq \frac{B}{\sqrt{L}}$} with a constant $B$, achieve \mbox{$C_\mathrm{DSSS}< C_\mathrm{AWGN}$} in the limit \mbox{$W\rightarrow\infty$} if \mbox{$\frac{L}{W}\rightarrow \alpha$} and \mbox{$\alpha>0$}. 
This bound holds whether the receiver knows the path delays or it does not.
\end{Converse}
\end{Theorem}
The proof is presented in Section~\ref{sec:DSSS}.
\begin{Theorem}\label{th:PPM_dknown}
PPM systems with duty cycle, where the receiver knows the path delays, achieve \mbox{$C_\mathrm{PPM}\rightarrow C_\mathrm{AWGN}$} as \mbox{$W\rightarrow\infty$}
if \mbox{$\frac{L}{\log{W}}\rightarrow 0$} and the path gains \mbox{$\left\{G_l\right\}_{l=1}^L$} satisfy
(A)~\mbox{$\max_{1 \le i \le L} |G_i| \rightarrow 0$} in probability as \mbox{$L \rightarrow \infty$}, 
and (B)~\mbox{$ E_G = \sum_{i=1}^L G_i^2 \rightarrow 1 $ in probability as $L \rightarrow \infty$}.
Note that if the gains are Gaussian IID with zero mean then the above condition holds.
\begin{Converse}
PPM systems with a non-vanishing symbol time, transmitting over a channel with Gaussian path gains that are unknown to the receiver, achieve \mbox{$C_\mathrm{PPM}\rightarrow0$} as \mbox{$W \rightarrow\infty$} if \mbox{$\frac{L}{\log W}\rightarrow \infty $}.
This result holds whether the receiver knows the path delays or it does not.
\end{Converse}
\end{Theorem}
The proof is presented in Section~\ref{sec:ppm}.

\section{Discussion}
\label{sec:discussion}

This section presents bounds on the data rates of direct sequence spread spectrum and PPM systems for different channels, computed in Sections~\ref{sec:penalty},~\ref{sec:DSSS}, and~\ref{sec:ppm}.
The channel and system parameters were chosen to represent a realistic low SNR ultra wide band system.
For the figures with fixed bandwidth we use:
Bandwidth $W$=20~GHz, $\frac{P}{N_0}$=53~dB (SNR=-50~dB at $W$=20~GHz), coherence period $T_c$=0.1~msec, delay spread $T_d$=200~nsec, PPM symbol time $T_s$=800~nsec with guard time of 200~nsec between symbols, and $B^2d=1$.
The constant $B$ is defined in the converse to Theorem~\ref{th:dsss}, it is used to characterize channel gains.
The constant $d$ is defined in~(\ref{eq:assumption2}) for pseudo-random chips; it equals 1 for IID chips.
For the figures with a fixed number of paths we use $L$=100.

\subsection{The Advantage of Duty Cycle}
\begin{figure}
\begin{center}
\leavevmode
\epsfysize=\figs \epsfbox{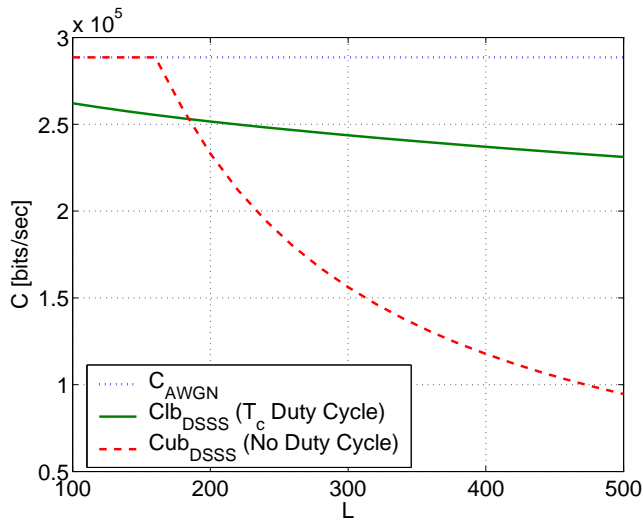}
\caption{
DSSS throughput bounds, the receiver does not know the channel gains nor the delays, vs.\ the number of channel paths.
This plot contrasts an upper bound on DSSS data rate, when duty cycle is not used (bottom graph, from~(\ref{eq:DSSS_ub})) with a lower bound on the data rate when coherence period duty cycle is used (top graph, from~(\ref{eq:DSSS_lb})). 
$C_\mathrm{AWGN}$ is shown for reference.
}
\label{fig:DSSScap}
\end{center}
\end{figure}

Figure~\ref{fig:DSSScap} shows the increase of data rate given by the usage of coherence period duty cycle, for DSSS systems.
The figure compares the upper bound on DSSS throughput, where duty cycle is not used (bottom graph) to the lower bound on throughput when optimal duty cycle is used.
Both bounds decrease as the number of paths $L$ increases because the channel uncertainty increases as $L$ increases, and so does the penalty on the data rate.

\begin{figure}
\begin{center}
\leavevmode
\epsfysize=\figs \epsfbox{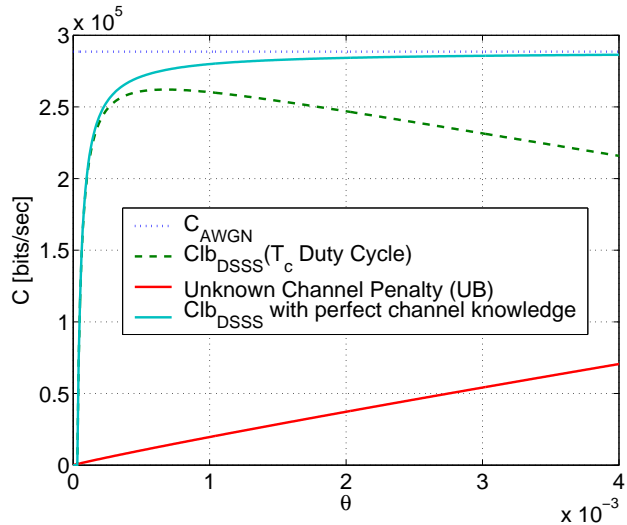}
\caption{
DSSS throughput lower bound vs.\ duty cycle parameter, the receiver does not know the channel path gains nor the delays.
The bottom curve shows an upper bound on the channel uncertainty penalty (calculated as the sum of~(\ref{eq:DSSS_lb_gainpenalty}) and~(\ref{eq:delaypenalty})).
The dashed curve shows a lower bound on the system throughput~(\ref{eq:DSSS_lb}), and the top (full) curve shows the throughput of a system with perfect channel knowledge at the receiver~(\ref{eq:DSSS_lb_1}).
$C_\mathrm{AWGN}$ is shown at the dotted curve for reference.
The system throughput (dashed curve) is the difference between the data rate of a system with perfect channel knowledge at the receiver (top curve) and the channel uncertainty penalty (bottom curve).
The throughput is maximized when the spectral efficiency is balanced against the channel uncertainty penalty.
}
\label{fig:DSSScaptheta}
\end{center}
\end{figure}

\subsection{The Duty Cycle Parameter}
Figure~\ref{fig:DSSScaptheta} shows the lower bound on the data rate of a direct sequence spread spectrum system for different duty cycle parameter values.
The bound is a difference between the data rate of a system with perfect channel knowledge at the receiver and the channel uncertainty penalty (gain penalty and delay penalty).
The data rate of a system with perfect channel knowledge equals the channel capacity ($C_\mathrm{AWGN}$) in the limit of infinite bandwidth, it is lower when the bandwidth is finite.

The channel uncertainty penalty is small for low values of duty cycle parameter, because the channel is used less often as $\theta$ decreases.
However, the data rate of a system with perfect channel knowledge is severely reduced if $\theta$ is too low.
In this case, transmission occurs with high energy per symbol, where the direct sequence spread spectrum modulation is no longer spectrally efficient, so the data rate with perfect channel knowledge is reduced.
Figure~\ref{fig:DSSScaptheta} shows that the duty cycle parameter must be chosen to balance the channel uncertainty penalty (that is large for large $\theta$) and the spectral efficiency of the selected modulation, that increases with $\theta$.

\subsection{Spectral Efficiency}
\begin{figure}
\begin{center}
\leavevmode
\epsfysize=\figs \epsfbox{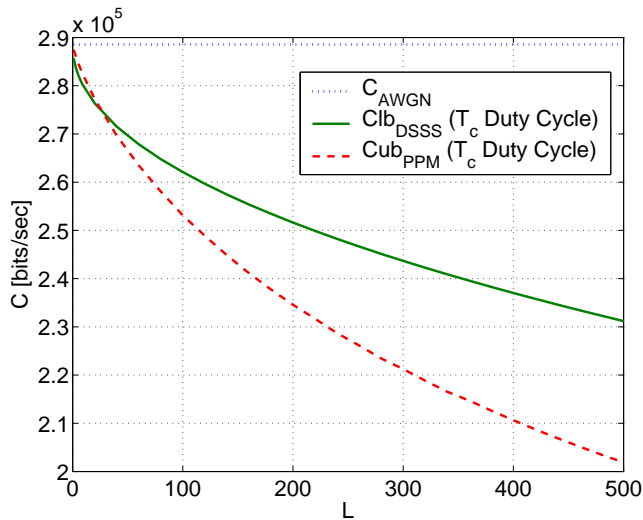}
\caption{
DSSS and PPM throughput bounds.
The DSSS lower bound~(\ref{eq:DSSS_lb}) is calculated without channel knowledge at the receiver. 
The PPM upper bound~(\ref{eq:PPM_ub}) is calculated with a receiver that knows the channel delays but not the gains, coherence period duty cycle is used.
$C_\mathrm{AWGN}$ is shown for reference.
}
\label{fig:DSSSPPMcap}
\end{center}
\end{figure}

\begin{figure}
\begin{center}
\leavevmode
\epsfysize=\figs \epsfbox{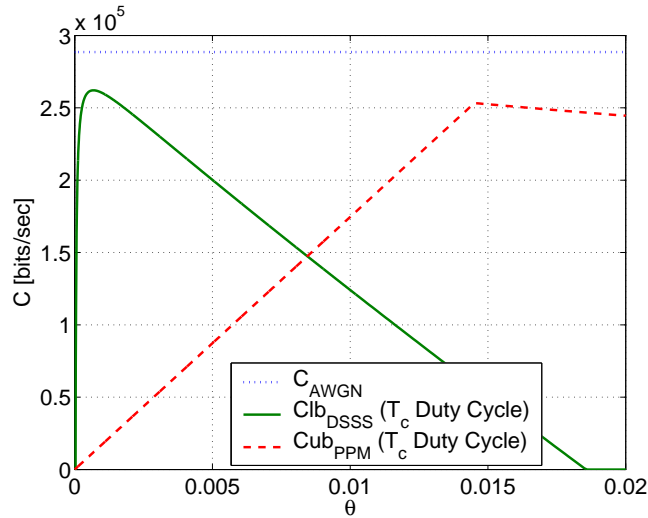}
\caption{
Throughput bounds vs.\ duty cycle parameter, the receiver knows the channel path delays but not the gains.
DSSS throughput lower bound~(\ref{eq:DSSS_lb}) is maximized at a lower duty cycle parameter than the PPM upper bound~(\ref{eq:PPM_ub}).
}
\label{fig:DSSSPPMcaptheta}
\end{center}
\end{figure}

Figure~\ref{fig:DSSSPPMcap} contrasts the achievable data rates of DSSS and PPM systems, when both use duty cycle on coherence periods with optimal duty cycle parameters.
Direct sequence spread spectrum achieves higher data rates because it has a higher spectral efficiency, thus it can pack more bits into each transmission period of length $T_c$. 
By packing bits efficiently a DSSS system is able to use a small duty cycle parameter. 
In contrast, PPM is less efficient in packing bits into its transmission periods, and is thus forced to transmit more often (it has a larger duty cycle parameter, Figure~\ref{fig:DSSSPPMcaptheta}).
The PPM system is therefore forced to handle a larger number of channel realizations (per unit time), so it suffers a higher penalty for estimating the channel parameters.

Spectral efficiency, in units of $\left[\frac{\mathrm{bits}}{\mathrm{sec}\,\mathrm{Hz}}\right]$, measures the number of bits that can be communicated per unit time per unit bandwidth.
The number of bits per DSSS symbol depends linearly on the bandwidth, thus its spectral efficiency does not depend on the bandwidth.
The number of bits per PPM symbol (with a fixed symbol time) depends logarithmically on the bandwidth, because PPM is an orthogonal modulation.
Thus, the PPM spectral efficiency depends on the bandwidth via $\frac{\log W}{W}$ and is much lower than the DSSS spectral efficiency if the bandwidth is large.

\section{Channel Uncertainty Penalty}
\label{sec:penalty}

The mutual information between the transmitted and the received signals is decomposed into the mutual information where the channel is known and a penalty term:
\begin{equation*}
I(X;Y)=I(X;Y|D,G)-I(X;D,G|Y)
\end{equation*}
where $I(X;D,G|Y)$ is the channel uncertainty penalty.
This can be decomposed:
\begin{equation*}
I(X;D,G|Y)=I(X;D|Y)+I(X;G|D,Y)
\end{equation*}
where $I(X;D|Y)$ is the {\em delay uncertainty penalty} and $I(X;G|D,Y)$ is the gain uncertainty penalty when the delays are known.

The delay uncertainty penalty is upper bounded by the entropy of the path delays.
The $L$ path delays are uniformly distributed over the $\binom{\lfloor WT_d\rfloor}{L}$ possibilities of combinations of $L$ values out of $WT_d$ positions spanning the delay spread.
\begin{equation}
I(X;D|Y)\leq H(D)\leq \frac{\theta L}{T_c}\log_2\left(WT_c\right)\ \ \ \left[\frac{\mathrm{bits}}{\mathrm{sec}}\right]  \label{eq:delaypenalty}
\end{equation}
and the gain uncertainty penalty when the delays are known is upper bounded (for DSSS signaling) by
\begin{equation*}
I(X;G|Y,D)\leq I(G;Y|X,D)\leq\frac{\theta L}{2T_c}\log_2\left(1+\frac{\mathcal{E}}{L}\right)\ \ \ \left[\frac{\mathrm{bits}}{\mathrm{sec}}\right]
\end{equation*}
see~(\ref{eq:DSSSgainpenalty}) for the derivation.



\section{Spread Spectrum Bounds}
\label{sec:DSSS}

\subsection{When is the Channel Capacity Achieved?}

We start with a result in the case of known path delays (Theorem~\ref{th:dsss} Part~\ref{prt:dsss_dknown}) that shows that the channel capacity is achieved if the number of paths is sub-linear with the bandwidth.
A second result is then given for the case of unknown channel, (Theorem~\ref{th:dsss} Part~\ref{prt:dsss_dunknown}), that shows that the channel capacity is still achievable, but at simpler environment, with a smaller number of channel paths.

\subsubsection{Path Delays Known to Receiver}
\label{sec:DSSS_lb_dknown}

\setcounter{Part}{1}
\addtocounter{Part}{-1}
\setcounter{tempCounter}{\value{Theorem}}
\setcounter{Theorem}{1}
\addtocounter{Theorem}{-1}
\begin{Theorem_nonum}
\begin{Part}
Direct sequence spread spectrum systems with duty cycle, where the receiver knows the path delays, achieve \mbox{$C_\mathrm{DSSS}\rightarrow C_\mathrm{AWGN}$} as \mbox{$W\rightarrow\infty$} if \mbox{$\frac{L}{W}\rightarrow 0$}.
\end{Part}
\end{Theorem_nonum}
\setcounter{Theorem}{\value{tempCounter}}

\begin{proof}
The proof is based on a lower bound on the mutual information.
\begin{Proposition}
\label{prop:dsss_dknown_lb}
DSSS systems where the receiver knows the channel path delays achieve
\begin{eqnarray}
&& I(X;Y|D)\ [\mathrm{b/s}] \geq C_\mathrm{AWGN} 
\label{eq:DSSS_lb_knownDelay} \\
& & - 
\min_{0<\theta\leq 1}\left\{
\frac{\theta{L}}{2T_c}\log_2\left( 1+\frac{P}{N_0}\frac{T_c}{\theta{L}}\right)
+\frac{3P^2 }{N_0^2\theta W} \log_2 e
\right\}
\nonumber
\end{eqnarray}
\end{Proposition}

{\em Discussion of Proposition \theProposition: }
The channel uncertainty penalty (due to path gains) has two parts, the first 
\begin{equation}
\frac{\theta{L}}{2T_c}\log_2\left( 1+\frac{P}{N_0}\frac{T_c}{\theta{L}}\right) \label{eq:DSSS_lb_gainpenalty}
\end{equation}
is the penalty due to the unknown gains, it increases as the number of paths ${L}$ or the duty cycle parameter $\theta$ increase.

The second part of the penalty 
\[
\frac{3P^2 }{N_0^2\theta W} \log_2 e
\]
is due to the limitation on spectral efficiency of the spread spectrum modulation.
It penalizes the system for using a too small duty cycle parameter, where the system concentrates too much energy on each transmitted symbol.
Mathematically, this term is the quadratic term in the series approximating the mutual information, that is logarithmic. 
The first (linear) term in this series equals $C_\mathrm{AWGN}$ in the limit.
The balance between the two penalties is shown in Figure~\ref{fig:DSSScaptheta}.

Looking at the limit of infinite bandwidth, the data rate converges to the AWGN capacity if \mbox{$\theta W \rightarrow \infty$} and \mbox{$\theta L\rightarrow 0$}.
If $L$ is sub--linear is $W$, these two requirements can be met simultaneously, that is, there exists a duty cycle parameter $\theta$, that depends on the bandwidth, such that $\theta L\rightarrow0$ and $\theta W\rightarrow\infty$.
Thus, the proof of Theorem~\ref{th:dsss} Part~\ref{prt:dsss_dknown} follows from Proposition~\ref{prop:dsss_dknown_lb}.

{\em Proof of Proposition \theProposition:}
The proof of~(\ref{eq:DSSS_lb_knownDelay}) follows Theorem~3 of~\cite{telatar_2000}, with a real channel instead of the complex channel used there.
We start from
\begin{equation*}
I(X;Y|D)=I(X;Y|G,D)-I(X;G|Y,D)
\end{equation*}
where the second summand is the gain uncertainty penalty with known delays.
By lower bounding this penalty:
\begin{equation}
I(X;G|Y,D)=I(G;Y|X,D)-I(G;Y|D)\leq I(Y;X|D)
\end{equation}
we get
\begin{equation}
I(X;Y|D)\geq I(Y;X|G,D)-I(Y;G|X,D) \label{eq:IDSSS_XYD}
\end{equation}

The first part of~(\ref{eq:IDSSS_XYD}):
\[
I(Y;X|G,D) =\frac{1}{2} E_{G,D} \log \det\left( I+\frac{\mathcal{E}}{WT_c} AA^\star\right)
\]
where $A$ is a \mbox{$K_c\times K_c$} matrix, \mbox{$A_{im} =G_l$} if $m=(i-D_l)$ and zero otherwise, and \mbox{$\mathcal{E}=\frac{2PT_c}{N_0\theta}$}.
The eigenvalues of \mbox{$AA^\star$} are \mbox{$\left| F\left(\frac{k}{K_c}\right)\right|^2$}, \mbox{$k=0,1,\dots, K_c-1$}, and
\[
F(f)=\sum_{l=1}^L G_l \exp(2\pi j D_l f ) 
\]
For large $L$ ($\,\gg 1$) $F(f)$ is complex Gaussian with independent real and imaginary parts that may have different variances (for small $f$), so \mbox{$E_G \left| F(f) \right|^2=1$} and \mbox{$E_G \left| F(f) \right|^4\leq3$}.
\begin{eqnarray*}
I(Y;X|G,D) & = &
\frac{1}{2}E_{G,D} \\
&& \left[ \sum_{k=0}^{WT_c-1} \log \left(1+\frac{\mathcal{E}}{WT_c} \left| F\left(\frac{k}{WT_c}\right)\right|^2\right)\right] \\
& \geq& 
\frac{\log e}{2}E_{G,D} \left[ \sum_{k=0}^{WT_c-1} \frac{\mathcal{E}}{WT_c} \left| F\left(\frac{k}{WT_c}\right)\right|^2 \right. \\
&&\left. -\frac{1}{2} \frac{\mathcal{E}^2}{W^2T_c^2} \left| F\left(\frac{k}{WT_c}\right)\right|^4 \right] \\
& \geq & 
\frac{\mathcal{E}}{2}\log e-\frac{3\mathcal{E}^2}{4WT_c} \log e \\
& = &
\frac{PT_c}{N_0\theta}\log e-\frac{3P^2T_c}{N_0^2\theta^2W} \log e  
\end{eqnarray*}
and In [bits/sec] units:
\begin{eqnarray}
I(Y;X|G,D) & \geq &
\frac{P}{N_0}\log_2 e-\frac{3P^2}{N_0^2\theta W} \log_2 e  
\label{eq:DSSS_lb_1}
\end{eqnarray}
where we used \mbox{$\log(1+x)\geq \left(x-\frac{x^2}{2}\right)\log e$}, that is valid for $x\geq0$.

The second part of~(\ref{eq:IDSSS_XYD}):
\[
I(Y;G|X,D) \leq\frac{1}{2}E_{X,D} \log \det\left( I+\frac{\mathcal{E}}{WT_c} B\Lambda B^\star\right)
\]
where \mbox{$B_{im}=X_{(i-m)}$} and \mbox{$\Lambda=\frac{1}{L}I$}.  
The upper bound is tight for Gaussian channel gains.
Following~\cite{telatar_2000} we get an upper bound
\begin{equation}
I(Y;G|X,D)\leq \frac{L}{2}\log\left(1+\frac{\mathcal{E}}{L}\right) \label{eq:DSSSgainpenalty}
\end{equation}

Rewriting~(\ref{eq:IDSSS_XYD}):
\begin{eqnarray*}
I(X;Y|D)& \geq & I(Y;X|G,D)-I(Y;G|X,D) \\
& \geq & 
\frac{\mathcal{E}}{2}\log e-\frac{3\mathcal{E}^2}{WT_c} \log e 
-\frac{L}{2}\log\left(1+\frac{\mathcal{E}}{L}\right)
\end{eqnarray*}

in [bits/sec]:
\begin{eqnarray*}
I(X;Y|D)\ [\mathrm{b/s}] & = & \frac{\theta I(X;Y|D)}{T_c} \\
& \geq &
\frac{\theta\mathcal{E}}{2T_c}\log_2 e -\frac{\theta{L}}{2T_c}\log_2\left( 1+\frac{\mathcal{E}}{{L}}\right) \\
&& -\frac{3\theta\mathcal{E}^2}{4K_cT_c}\log_2 e  \\
& = &
\frac{P}{N_0}\log_2 e -\frac{\theta{L}}{2T_c}\log_2\left( 1+\frac{P}{N_0}\frac{2T_c}{\theta{L}}\right) \\
&& -\frac{3P^2 }{N_0^2\theta W} \log_2 e 
\end{eqnarray*}
The bound is valid for any $\theta$, and we choose its maximal value:
\begin{eqnarray*}
&& I(X;Y|D)\ [\mathrm{b/s}] \geq \\
&&  \max_{0<\theta\leq 1}\left\{
\frac{P}{N_0}\log_2 e \right.\\
&& \left.-\frac{\theta{L}}{2T_c}\log_2\left( 1+\frac{P}{N_0}\frac{2T_c}{\theta{L}}\right)  -\frac{P^2 }{N_0^2\theta W} \log_2 e
\right\} \nonumber \\
& = &
C_\mathrm{AWGN}
 \\
&& -
\min_{0<\theta\leq 1}\left\{
\frac{\theta{L}}{2T_c}\log_2\left( 1+\frac{P}{N_0}\frac{2T_c}{\theta{L}}\right)
+\frac{P^2 }{N_0^2\theta W} \log_2 e
\right\}
\end{eqnarray*}
\end{proof}

\subsubsection{Path Delays Unknown to Receiver}
\label{sec:DSSS_lb_dunknown}

\setcounter{Part}{2}
\addtocounter{Part}{-1}
\setcounter{tempCounter}{\value{Theorem}}
\setcounter{Theorem}{1}
\addtocounter{Theorem}{-1}
\begin{Theorem_nonum}
\begin{Part}
DSSS systems with duty cycle achieve \mbox{$C_\mathrm{DSSS}\rightarrow C_\mathrm{AWGN}$} as \mbox{$W\rightarrow\infty$} if \mbox{$\frac{L\log{W}}{W}\rightarrow 0$}.
\end{Part}
\end{Theorem_nonum}
\setcounter{Theorem}{\value{tempCounter}}

\begin{proof}
The proof is based on Proposition~\ref{prop:dsss_dknown_lb} and equation~(\ref{eq:delaypenalty}) that relates the mutual information in the case of channel knowledge of the path delays with the mutual information in the general case.
With no channel knowledge at the receiver (path delays and gains unknown), we get:
\begin{eqnarray}
I(X;Y)\ [\mathrm{b/s}] & \geq & 
C_\mathrm{AWGN} 
\label{eq:DSSS_lb} \\
&&  -
\min_{0<\theta\leq 1}\left\{
\frac{\theta{L}}{T_c}\log_2\left( 1+\frac{P}{N_0}\frac{T_c}{\theta{L}}\right) \right. \nonumber \\
&& \left. +\frac{3P^2 }{N_0^2\theta W} \log_2 e 
+\frac{\theta L}{T_c}\log_2\left(WT_d\right)
\right\}
\nonumber
\end{eqnarray}
The third penalty term describes the penalty due to path delays, from~(\ref{eq:delaypenalty}).
This term is a bound on the penalty, that depends linearly on the number of path delays per unit time.

At the limit of infinite bandwidth the bound equals the AWGN capacity if
\begin{itemize}
\item \mbox{$\theta W \rightarrow \infty$} 
\item \mbox{$\theta L\rightarrow 0$} 
\item \mbox{$\theta L\log W\rightarrow 0$}
\end{itemize}
The second condition may be dropped, as the third is stronger. 
These conditions can be met simultaneously, that is, there exists a duty cycle parameter $\theta$ that depends on the bandwidth and satisfies the conditions, if $L\log W$ is sub--linear in $W$, namely \mbox{$\frac{L\log W}{W}\rightarrow0$}.
\end{proof}

\subsection{When is the Channel Capacity Not Achieved?}
\label{sec:DSSS_ub}

An additional assumption on gains is used in this section: the gains are uniformly upper bounded by 
\mbox{$|G_l|\leq \frac{B}{\sqrt{L}}$}, this is a technical condition that follows~\cite{telatar_2000}.
\setcounter{tempCounter}{\value{Theorem}}
\setcounter{Theorem}{1}
\addtocounter{Theorem}{-1}
\begin{Theorem_converse}
DSSS systems with duty cycle where the path gains are unknown to the receiver and uniformly bounded by \mbox{$|G_l|\leq \frac{B}{\sqrt{L}}$} with a constant $B$, achieve \mbox{$C_\mathrm{DSSS}< C_\mathrm{AWGN}$} in the limit \mbox{$W\rightarrow\infty$} if \mbox{$\frac{L}{W}\rightarrow \alpha$} and \mbox{$\alpha>0$}. 
This bound holds whether the receiver knows the path delays or it does not.
\end{Theorem_converse}
\setcounter{Theorem}{\value{tempCounter}}

\begin{proof}
We first note that the mutual information in a system where the receiver knows the path delays upper bounds the mutual information in the general case:
\[
I(X;Y)=I(X;Y|D)-I(D;X|Y)\leq I(X;Y|D)
\]
So we only need to prove the theorem regarding the conditional mutual information, where the receiver knows the path delays. 

The proof is based on the following upper bound on the mutual information.
\begin{Proposition}
\label{prop:dsss_dknown_ub}
DSSS systems with duty cycle parameter $\theta$ achieve
\begin{eqnarray}
I(X;Y|D)\ [\mathrm{b/s}] & \leq & 
W\theta \log_2\left(1+\frac{P}{N_0W\theta}\right) 
\label{eq:DSSS_ub1}
\end{eqnarray}
If the duty cycle is chosen so that \mbox{$\theta L\rightarrow\infty$} as \mbox{$W\rightarrow\infty$}, then a second upper bound holds in the limit:
\begin{eqnarray}
I(X;Y|D)\ [\mathrm{b/s}] & \leq &
C_\mathrm{AWGN}\frac{2T_dB^2d}{T_c}
\label{eq:DSSS_ub2}
\end{eqnarray}
$d$ is defined in~(\ref{eq:assumption2}) for pseudo-random chips.
For IID chips, the bound holds with $d=1$.
\end{Proposition}

{\em Discussion of Proposition \theProposition:}
We now look at two different possibilities regarding the duty cycle parameter $\theta$:
\begin{itemize}
\item \mbox{$\theta W<\infty$}. In this case the bound~(\ref{eq:DSSS_ub1}) is strictly lower than $C_\mathrm{AWGN}$.
\item \mbox{$\theta W\rightarrow\infty$} (as \mbox{$W\rightarrow\infty$}). Using our assumption on the number of channel paths we get \mbox{$\theta L\rightarrow\infty$}, so the second bound~(\ref{eq:DSSS_ub2}) becomes relevant.
In situations where \mbox{$T_d\ll T_c$}, this bound is significantly lower than the AWGN capacity.
\end{itemize}

If the number of paths is sub--linear in $W$, the duty cycle can be chosen so that \mbox{$\theta W\rightarrow\infty$} and \mbox{$\theta L\rightarrow0$} and the bounds in~(\ref{eq:DSSS_ub2}) become irrelevant.
In this case the upper bound converges to $C_\mathrm{AWGN}$ in the limit of infinite bandwidth.

To summarize the behavior of the bound in the limit, in the case of a linear increase of the number of paths with the bandwidth, the upper bound is lower than the AWGN capacity in normal operating conditions ($T_d\ll T_c$).
The upper bound equals $C_\mathrm{AWGN}$ in the case of a sub-linear increase of the number of paths with the bandwidth.

{\em Proof of Proposition \theProposition}
We start with a simple bound:
\begin{eqnarray*}
I(X;Y|D)\ [\mathrm{b/s}] & \leq  & 
 W\theta \log_2\left(1+\frac{P}{N_0W\theta}\right) 
\end{eqnarray*}
This is the capacity of an AWGN channel used with duty cycle $\theta$.
It upper bounds the data rate for systems with channel knowledge at the receiver.
In order to achieve the capacity at the limit of infinite bandwidth, the duty cycle parameter must be chosen so that \mbox{$\theta W\rightarrow\infty$}.

To prove~(\ref{eq:DSSS_ub2}) we follow Theorem~2 of~\cite{telatar_2000}, that gives an upper bound on the mutual information.
\[
 I(X;Y|D) \leq E_G \log\left( E_H \exp \left[ 2\mathcal{E} \sum_{l=1}^L H_l G_l \right]\right) 
\]
\begin{equation}
 +   2\mathcal{E} \frac{B^2}{L}  E_D\left[ \sum_{l=1}^L \sum_{m=1}^L E_X\left| C(D_l, D_m) -\delta_{lm}\right|\right]\log_2e \label{eq:Ib}
\end{equation}
where $G$ and $H$ are identically distributed and independent, \mbox{$|H_l|\leq \frac{B}{\sqrt{L}}$}. 

The first part of~(\ref{eq:Ib}):
\begin{eqnarray*}
&& \log E_H\exp \left[ 2\mathcal{E} \sum_{l=1}^L H_l G_l \right] 
 \\
&& = \log E_{|H|} E_\psi \exp \left[ 2\mathcal{E} \mathrm{Re} \left\{ \sum_{l=1}^L e^{-j\psi_l} |H_l| G_l\right\} \right]
\end{eqnarray*}
The phases \mbox{$\left\{\psi_l\right\}$} equal $0$ or $\pi$ with probability $1/2$.
\begin{eqnarray*}
E_{\psi} \exp(ae^{j\psi_l})=\frac{e^a+e^{-a}}{2}\approx 1+\frac{a^2}{2} \ \ \ \ \ \ \mathrm{for}\ a\ll1
\end{eqnarray*}
\begin{eqnarray*}
E_{|H|} E_\psi \exp \left[ 2\mathcal{E} \sum_{l=1}^L e^{-j\psi_l} |H_l| G_l \right]  \\
\leq 
E_{|H|} \prod_{l=1}^L \left( 1+ 2\mathcal{E}^2 |H_l|^2|G_l|^2 \right)
\end{eqnarray*}
and the condition is \mbox{$\frac{2\mathcal{E}B^2}{L}\ll1$}, which holds if \mbox{$\theta L\rightarrow\infty$}.
\begin{eqnarray*}
E_G \log\left( E_H \exp \left[ 2\mathcal{E} \sum_{l=1}^L H_l G_l \right]\right) 
\\
\leq
E_G \log E_{|H|} \prod_{l=1}^L \left( 1+2\mathcal{E}^2 |H_l|^2|G_l|^2 \right)
\end{eqnarray*}
Using Jensen's inequality:
\begin{eqnarray*}
&&  E_G \log\left( E_H \exp \left[ 2\mathcal{E} \sum_{l=1}^L H_l G_l \right]\right)  \\
&& \leq 
\sum _{l=1}^L \log \left( 1 +E_{|G_l|,|H_l|} 2\mathcal{E}^2 |H_l|^2|G_l|^2 \right) \\
&& =  
\sum _{l=1}^L \log \left( 1 +\frac{2\mathcal{E}^2}{L^2} \right) \\
&& =  
L \log \left( 1 +\frac{2\mathcal{E}^2}{L^2} \right)
\end{eqnarray*}
with \mbox{$\mathcal{E}=\frac{2PT_c}{N_0\theta}$}:
\begin{eqnarray*}
&&  E_G \log\left( E_H \exp \left[ 2\mathcal{E} \mathrm{Re}\left\{ \sum_{l=1}^L H_l G_l\right\} \right]\right)  \nonumber \\
&& \leq 
L \log \left( 1 +\frac{8P^2T_c^2}{N_0^2\theta^2L^2} \right) \nonumber \\
&& \leq 
 \frac{8P^2T_c^2}{N_0^2\theta^2L}\log e  
\end{eqnarray*}

The second part of~(\ref{eq:Ib}):
\begin{itemize}
\item For IID chips:
Using~(\ref{eq:dsss_iid_cond}) 
\begin{eqnarray*}
&& 2\mathcal{E} \frac{B^2}{L} E_D\left[ \sum_{l=1}^L \sum_{m=1}^L E_X\left| C(D_l, D_m) -\delta_{lm}\right|\right] \log_2e \nonumber \\
&& \leq 
2\mathcal{E} \frac{B^2}{L} E_D
\left( \frac{3L}{\sqrt{K_c}} +\frac{L^2-L}{K_c} \right) 
 \log_2e \\
&& = 
2\mathcal{E} B^2
\left( \frac{3}{\sqrt{K_c}} +\frac{L-1}{K_c} \right)
 \log_2e \\
&& \approx 
\frac{2\mathcal{E} B^2 L}{K_c}
 \log_2e \nonumber \ \ \ \ \ \ \mathrm{for}\ L\gg 1  \\
&& \leq 
\frac{4PT_d}{N_0\theta} B^2\log_2e 
\end{eqnarray*}
The last inequality follows from \mbox{$L\leq T_dW$}.
\item For pseudo-random chips:
\begin{eqnarray}
&& 2\mathcal{E} \frac{B^2}{L} E_D\left[ \sum_{l=1}^L \sum_{m=1}^L E_X\left| C(D_l, D_m) -\delta_{lm}\right|\right] \log_2e \nonumber \\
&& \leq 
2\mathcal{E} B^2 L \frac{d}{K_c}\log_2e \nonumber \\
&& = 
\frac{4PL}{N_0\theta W} B^2d\log_2e \nonumber \\
&& \leq 
\frac{4PT_d}{N_0\theta} B^2d\log_2e \label{eq:Ib_2}
\end{eqnarray}
\end{itemize}
so~(\ref{eq:Ib_2}) is valid in both cases of input signals, and for IID chips we take $d=1$.

Putting the two parts back into~(\ref{eq:Ib}):
\[
I(X;Y|D) \leq 
\frac{8P^2T_c^2}{N_0^2\theta^2L} \log e  
+
\frac{4PT_d}{N_0\theta } B^2d\log_2e 
\]
In units of [bits/sec]:
\begin{equation}
I(X;Y|D)\ [\mathrm{b/s}] \leq 
\frac{8P^2T_c}{N_0^2\theta L} \log_2 e  
+
\frac{4PT_d}{N_0T_c} B^2d\log_2e \label{eq:DSSS_ub}
\end{equation}
Using \mbox{$\theta L\rightarrow\infty$} we get 
\[
I(X;Y|D)\ [\mathrm{b/s}] \leq 
\frac{4PT_d}{N_0T_c} B^2d\log_2e 
=C_\mathrm{AWGN}\frac{2T_dB^2d}{T_c} 
\]
\end{proof}

\section{PPM Bounds}
\label{sec:ppm}

\subsection{When is the Channel Capacity Achieved? (Path Delays Known to Receiver)}
\label{sec:PPM_dknown}


\setcounter{tempCounter}{\value{Theorem}}
\setcounter{Theorem}{2}
\addtocounter{Theorem}{-1}
\begin{Theorem_nonum}
PPM systems with duty cycle, where the receiver knows the path delays, achieve \mbox{$C_\mathrm{PPM}\rightarrow \CAWGN$} as \mbox{$W\rightarrow\infty$}
if \mbox{$\frac{L}{\log{W}}\rightarrow 0$} and the path gains \mbox{$\left\{G_l\right\}_{l=1}^L$} satisfy
(A)~\mbox{$\max_{1 \le i \le L} |G_i| \rightarrow 0$} in probability as \mbox{$L \rightarrow \infty$}, 
and (B)~\mbox{$ E_G = \sum_{i=1}^L G_i^2 \rightarrow 1 $ in probability as $L \rightarrow \infty$}.
Note that if the gains are Gaussian then the above conditions hold.
\end{Theorem_nonum}
\setcounter{Theorem}{\value{tempCounter}}

\begin{proof}
We start by breaking the mutual information in two parts:
\begin{equation}
I(X;Y|D)\geq I(X;Y|G,D)-I(Y;G|X,D) \label{eq:ppmlb_start}
\end{equation}
The maximal data rate achievable by systems that use the PPM signals defined in Section~\ref{sec:PPM_signal} is the maximum of~(\ref{eq:ppmlb_start}) over the duty cycle parameter $\theta$.
\begin{eqnarray}
C_\mathrm{PPM}& =& \max_\theta I(X;Y|D) \nonumber \\
& \geq & \max_\theta\left[ I(X;Y|G,D)-I(Y;G|X,D)\right]
\label{eq:PPM_cstart}
\end{eqnarray}

The first part of~(\ref{eq:PPM_cstart}) describes the throughput of a system that knows the channel perfectly.
Section~\ref{sec:IXY_GD} shows that it approaches $\CAWGN$ if the duty cycle parameter $\theta$ is chosen appropriately.
This result is shown by demonstrating that the probability of error diminishes as the bandwidth increases while the data rate is as close to $\CAWGN$ as desired.
Our analysis shows not only that the probability of error diminishes, it also demonstrates a specific reception technique, showing achievability of the capacity in the limit.
Section~\ref{sec:IYG_XD} calculates the penalty on  the system for its channel gain knowledge and shows that it diminishes for our choice of duty cycle parameter $\theta$.

The receiver we use in the analysis of probability of error is based on a matched filter, it is derived from the optimal (maximum likelihood) receiver in the case of a channel with a single path.
PPM signals are composed of {\em orthogonal} symbols. 
When a PPM signal is transmitted over an impulsive (single path) channel, the orthogonality of the symbols is maintained at the receiver side.

Considering a multipath channel, the received symbol values are no longer orthogonal, so the matched filter receiver is no longer optimal.
The non--orthogonality of the received symbols has an adverse effect on receiver performance. 
As the number of channel paths increases, the received symbols become increasingly non--orthogonal, and the receiver performance is degraded.
The matched filter receiver can sustain a growth of the number of channel paths (as the bandwidth increases), but this growth must not be too rapid.
To put it more formally, our system achieves the channel capacity in the limit of infinite bandwidth, if the number of paths obeys \mbox{$\frac{L^6}{W}\rightarrow0$}.

For each possible transmitted symbol value, the receiver matches the signal with a filter that has $L$ fingers at the right delays (Figure~\ref{fig:PPM_overlapsig}).
\begin{figure}
\begin{center}
\leavevmode
\epsfysize=\figs \epsfbox{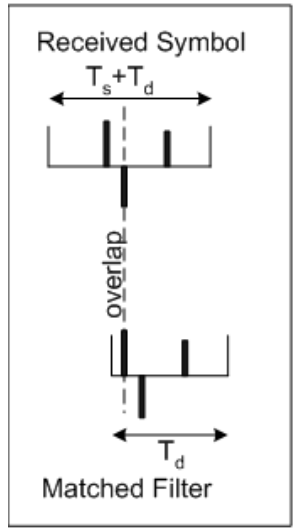}
\caption{
A received symbol and the matched filter showing a partial overlap.
}
\label{fig:PPM_overlapsig}
\end{center}
\end{figure}
The receiver is essentially of the rate type, it constitutes a matched filter to the received signal, that depends on the (known) channel.
The receiver uses a threshold parameter \mbox{$A=\alpha \sqrt{\mathcal{E}/N}$} for deciding on the transmitted value.
If one output only is above $A$, the input is guessed according to this output.
If none of the outputs pass $A$, or there are two or more that do, an error is declared.

We calculate an upper bound on the probability of error of this system, and show that it converges to zero as the bandwidth increases, if the number of channel paths does not increase too rapidly, namely \mbox{$\frac{L^6}{W}\rightarrow 0$}, and the duty cycle parameter is chosen properly.

The second part of~(\ref{eq:PPM_cstart}) describes a penalty due to unknown path gains, it is analyzed separately in Section~\ref{sec:IYG_XD}, the upper bound calculated there does not depend on the coding used by the transmitter, and it diminishes as the bandwidth increases for our choice of duty cycle parameter. 

We summarize here the conditions for convergence of~(\ref{eq:PPM_cstart}), to bring the conclusion of the following lengthy calculation:
The system uses duty cycle with parameter $\theta$ over coherence periods;
 The first part of~(\ref{eq:PPM_cstart}) converges to $\CAWGN$ in the limit of infinite bandwidth, if the following conditions take place (end of Section~\ref{sec:IXY_GD}):
 \begin{itemize}
\item \mbox{$\theta \log L\rightarrow 0$}
\item $\frac{L^6}{W}\rightarrow0$
\item \mbox{$\theta\log W \sim \mathrm{const}$}
 \end{itemize}
 The second part of~(\ref{eq:PPM_cstart}) contains the penalty for channel (gain) uncertainty; it converges to zero if \mbox{$\theta L\rightarrow0$} (Section~\ref{sec:IYG_XD}).
These conditions can exist simultaneously if \mbox{$\frac{L}{\log W}\rightarrow0$}.

\subsubsection{Upper Bound on $I(Y;G|X,D)$}
\label{sec:IYG_XD}
The position of the signal fingers is known as $X$ and $D$ are known.
\begin{eqnarray}
I(Y;G|X,D) & = & h(Y|X,D)-h(Y|X,G,D) \label{eq:I1}
\end{eqnarray}
The first part of~(\ref{eq:I1}) is upper bounded by the differential entropy in the case of Gaussian path gains.
Considering the signals during a coherence period  ($N$ transmitted symbols), the discretized received signal is composed of $NM_r$ values (chips), where \mbox{$M_r=W(T_s+T_d)$}.
Given $X$ and $D$, it is known which chips contain signal and which contain only noise (Figure~\ref{fig:PPM_Rxsignal}).
\begin{figure}
\begin{center}
\leavevmode
\epsfxsize=\figswide \epsfbox{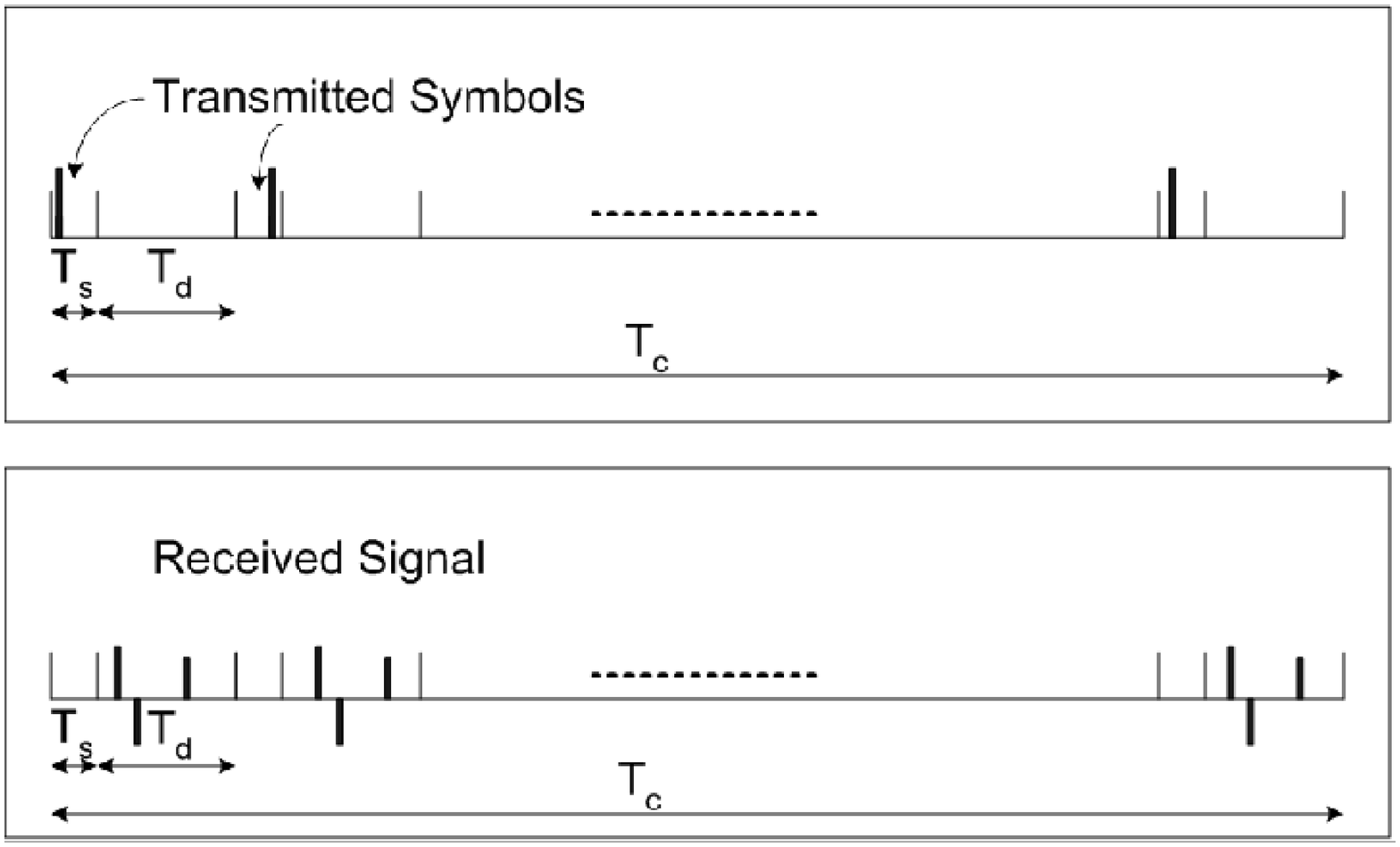}
\caption{
The transmitted PPM signal, with guard time of $T_d$ between symbols, and the received signal.
}
\label{fig:PPM_Rxsignal}
\end{center}
\end{figure}
The $NM_r$ received values are distributed as follows:
\begin{itemize}
\item \mbox{$N(M_r-L)$} are IID Gaussian \mbox{$\sim N(0,1)$}.
\item $NL$ values are divided into groups of size $N$.
Each group is independent of the other groups, it has zero mean and its correlation matrix is
\begin{eqnarray*}
\Lambda & = & \frac{2P\left(T_s+T_d\right)}{\theta N_0L}+\left(\begin{array}{ccc} 1  & & 0 \\ & \ddots & \\0 & & 1 \end{array}\right) 
\end{eqnarray*}
\end{itemize}
The differential entropy (in bits per coherence time) is bounded by
\[
h(Y|X,D)\leq\frac{N(M_r-L)}{2}\log_2(2\pi e)+\frac{L}{2}\log_2\left((2\pi e)^N\left|\Lambda\right|\right)
\]
The determinant $|\Lambda|$ is the product of the eigenvalues of
$\Lambda$: $1$ with multiplicity $N-1$ and
\mbox{$1+\frac{\mathcal{E}}{L}$} with multiplicity one
\[
|\Lambda|=1+\frac{\mathcal{E}}{L}
\]

The second part of~(\ref{eq:I1}) is given by
\[
h(Y|X,D,G)=\frac{NM_r}{2}\log_2\left(2\pi e\right)
\]
Combining both parts, and translating to units of \mbox{$\left[\mathrm{bits/sec}\right]$}:
\begin{eqnarray}
I(Y;G|X,D)\ [\mathrm{b/s}] & \leq &
\frac{\theta L}{2T_c} \log_2 \left|\Lambda\right| \nonumber \\
& = & 
\frac{\theta L}{2T_c} \log_2 \left(1+\frac{\mathcal{E}}{L}\right)  \nonumber \\
& = & 
\frac{\theta L}{2T_c} \log_2 \left(1+\frac{2P\left(T_s+T_d\right)N}{\theta N_0L}\right) \label{eq:I2} 
\end{eqnarray}
The bound~(\ref{eq:I2}) converges to zero as \mbox{$\theta L\rightarrow0$}.

\subsubsection{Lower Bound on $\max_\theta I(Y;X|G,D)$}
\label{sec:IXY_GD}
This bound holds if the path gains \mbox{$\left\{G_l\right\}_{l=1}^L$} satisfy
(A)~\mbox{$\max_{1 \le i \le L} |G_i| \rightarrow 0$} in probability as \mbox{$L \rightarrow \infty$}, 
and (B)~\mbox{$ E_G = \sum_{i=1}^L G_i^2 \rightarrow 1 $ in probability as $L \rightarrow \infty$}.
We first show that the Gaussian distribution \mbox{$G_l\sim N\left(0,1/L\right)$} satisfies these conditions.
Condition~(B) follows easily from the law of large numbers.
To prove that condition~(A) holds, we use the following well--known tail estimate for a standard normal $Z$ and any $x>0$:
\begin{equation}\label{ztails}
P(Z>x)\leq \frac{1}{\sqrt{2\pi}x} \exp\left(-x^2/2\right)
\end{equation}
Using \mbox{$\beta=2 \sqrt{\frac{\log L}{L}}$} we get
\begin{eqnarray*}
P\left(\max_{1\leq l\leq L}\left|G_l\right|>\beta\right)
& \leq & LP\left(\left|G_1\right|>\beta\right) \\
& \leq & LP\left(\left|Z\right|>\beta\sqrt{L}\right) \\
& \leq & \frac{L\sqrt{L}}{\sqrt{2\pi\log L}}\exp (-2 \log L) \\
& \rightarrow & 0 \ \ \ \mathrm{as}\ L\rightarrow\infty
\end{eqnarray*}
Clearly \mbox{$\beta \rightarrow 0$} as \mbox{$L\rightarrow\infty$}.

{\em Analysis of the Signals in the Receiver:}
For every symbol, the receiver calculates
\[
s_i=\sum_{j=1}^L G_j Y_{i+D_j-1} \ \ \ \ \ i=1,\dots, WT_s
\]
Assuming that $x_1$ was transmitted the desired output is Gaussian with
\begin{eqnarray*}
E[s_1] & = & \sqrt{\mathcal{E}/N}E_G \\
\sigma_{s_1}^2 & = & E_G
\end{eqnarray*}

There are up to $L^2-L$ Gaussian overlap terms that contain part of the signal (Figure~\ref{fig:PPM_overlapsig}).
Each of these overlap terms can be described by a set $\mathrm{OL}$ of pairs of path gains, each index $I_{\circ,\circ}$ (between 1 and $\lfloor T_dL\rfloor$) indicates a path in the channel response.

\[
\mathrm{OL}=\left\{\left(I_{1,1}, I_{1,2}\right), \dots, \left(I_{|\mathrm{OL}|,1},I_{|\mathrm{OL}|,2} \right)\right\}
\]
The number of terms in the set may vary in the range \mbox{$1\leq|\mathrm{OL}|\leq L-1$} and the indices take values between 1 and \mbox{$\lfloor T_dW\rfloor$}.
The pairs of indices are composed of two different indices, because the case where the filter position corresponds to the actual signal position is already accounted for in $s_1$. 

Given the overlap positions, or the set OL, the overlap terms are Gaussian with
\begin{eqnarray*}
E[s_\mathrm{overlap}|\mathrm{OL}] & = & \sum_{i=1}^{|\mathrm{OL}|}G_{I_{i,1}}G_{I_{i,2}}\sqrt{\mathcal{E}/N} \\
E[s_\mathrm{overlap}^2|\mathrm{OL}] & \leq & \sum_{i=1}^{|\mathrm{OL}|}G_{I_{i,1}}^2G_{I_{i,2}}^2\mathcal{E}/N+E_G
\end{eqnarray*}

Assuming a small number of paths, the probability that there are two or more overlaps (the set OK has more than one element)
converges to zero as the bandwidth increases to infinity, see the proof of this convergence in Section~\ref{sec:numofpaths}.
The assumption on the number of paths is satisfied if \mbox{$\frac{L^4}{W}\rightarrow 0$}.
Additional conditions on $L$, stemming from other parts of the proof, will require an even slower increase of $L$ with the bandwidth.

In addition, there are up to $WT_s-1$ Gaussian noise terms:
\begin{eqnarray*}
E[s_\mathrm{noise}] & = & 0 \\
E[s_\mathrm{noise}^2] & = & E_G
\end{eqnarray*}

For each possible transmitted symbol value, the receiver compares \mbox{$\left\{s_i\right\}_{i=1}^{WT_s}$} to a threshold \mbox{$A=\alpha \sqrt{\mathcal{E}/N}=\alpha\sqrt{\frac{2PT_c}{N_0\theta N}}$} where \mbox{$\alpha \in (0,1)$}.
If one output only is above $A$, the input is guessed according to this output.
If none of the outputs pass $A$, or there are two or more that do, an error is declared. \\

There are three types of error events, and the error probability is upper bounded using the union bound:
\begin{eqnarray}
P(\mathrm{error}) & \leq  & P(s_1\leq A)+ (L^2-L)P(s_\mathrm{overlap}\geq A) \nonumber \\
&& +(WT_s-1)P(s_\mathrm{noise}\geq A) \nonumber \\
& \leq  & P(s_1\leq A)+ L^2P(s_\mathrm{overlap}\geq A) \nonumber \\
&& +WT_sP(s_\mathrm{noise}\geq A) 
\label{eq:PeGun}
\end{eqnarray}
The first probability is bounded using the Chebyshev inequality, and the second and third using the normal tail estimate.

{\em First Error Event:}
Recall $s_1$ has expectation \mbox{$\sqrt{\mathcal{E}/N} E_G$} and variance $ E_G $.
From the Chebyshev inequality,
\begin{eqnarray*}
P(s_1\leq A) & \leq & \frac{\sigma_{s_1}^2}{\left(E[s_1]-A\right)^2} \\
& = & \frac{ E_G }{ (E_G - \alpha)^2 \mathcal{E}/N}
\end{eqnarray*}
Since $\alpha < 1$ and \mbox{$ E_G \rightarrow 1 $} in probability, for large $L$
the ratio \mbox{$E_G / (E_G - \alpha)^2 $} is bounded. Since \mbox{$\mathcal{E}/N \rightarrow \infty$}, the probability converges to 0.

{\em Second Error Event:}
The probability that an overlap term exceeds the threshold is expressed as a sum over the $L-1$ possibilities of the number of overlap positions:
\begin{eqnarray*}
&& L^2P ( s_\mathrm{overlap} \ge A ) \\
&& = L^2\sum_{i=1}^{L-1}P\left(\left|\mathrm{OL}\right|=i\right) P\left(s_\mathrm{overlap}\geq A|\left|\mathrm{OL}\right|=i\right)
\end{eqnarray*}
Section~\ref{sec:numofpaths} shows that if the number of paths is such that \mbox{$\frac{L^4}{W}\rightarrow0$}, then the probability of overlap at more than one position diminishes as \mbox{$W\rightarrow\infty$}.
\[
P\left(\left|\mathrm{OL}\right|>1\right)\rightarrow0
\]
In order to ensure that the overlap terms with more than one overlap position are insignificant in the calculation of the probability of error, we require \mbox{$\frac{L^6}{W}\rightarrow0$}, and then get in the limit of large bandwidth
\begin{eqnarray*}
L^2P ( s_\mathrm{overlap} \ge A ) & \rightarrow & L^2 P\left(s_\mathrm{overlap}\geq A|\left|\mathrm{OL}\right|=1\right)
\end{eqnarray*}
The condition \mbox{$|\mathrm{OL}|=1$} is omitted in the remainder of the calculation.

Recall that in the single overlap case
$ s_\mathrm{overlap} $ is normal with mean \mbox{$ \mu = G_l G_m \sqrt{\mathcal{E}/N} $} and variance \mbox{$ E_G = \sum G_i^2 $}. Hence \mbox{$ ( s_\mathrm{overlap} - \mu ) / \sqrt{E_G} $}
is a standard normal. \\
By assumption, \mbox{$ \max |G_i| \rightarrow 0 $} and \mbox{$ E_G \rightarrow 1 $} in 
probability, so for $L$ large we can assume \linebreak[4] \mbox{$ \mu \le \alpha / 2  \sqrt{\mathcal{E}/N} = A/2 $}
and $ E_G \le 4 $. Then
\begin{eqnarray*}
&& P ( s_\mathrm{overlap} \ge A ) \\
&& =  P ( ( s_\mathrm{overlap} - \mu ) / \sqrt{E_G} \ge ( A - \mu ) / \sqrt{E_G} ) \\
&& \le  P ( ( s_\mathrm{overlap} - \mu ) / \sqrt{E_G} \ge A / 4) = P\left(Z\geq A/4\right)
\end{eqnarray*}
where $Z$ stands for a standard normal.
Using the normal tail estimate~(\ref{ztails}), we obtain
\begin{eqnarray*}
L^2 P ( s_\mathrm{overlap} \ge A ) & \le & L^2 P ( Z \ge A / 4) \\
& \le & \exp ( 2 \ln L - A^2 / 32 - \ln A / 2 )
\end{eqnarray*}
In order to ensure convergence to zero of this probability, it is enough to have \mbox{$ (\ln L) / A^2 \rightarrow 0 $}.
Recalling \mbox{$A=\alpha \sqrt{\mathcal{E}/N}=\alpha\sqrt{\frac{2PT_c}{N_0\theta N}}$}, we get an equivalent condition: \mbox{$\theta \ln L \rightarrow 0 $}.

{\em Third Error Event:}
Recall $ s_\mathrm{noise} $ is normal with mean 0 and variance $E_G$.
The third probability in~(\ref{eq:PeGun}) is upper bounded using the normal tail estimate~(\ref{ztails})
for the standard normal \mbox{$Z = s_\mathrm{noise} / \sqrt{E_G}$}:
\[
P( s_\mathrm{noise} \geq A) \leq (2\pi)^{-1/2} (\sqrt{E_G}/A) \exp(-A^2/(2E_G))
\]
\[
WT_sP(s_\mathrm{noise}\geq A) \leq \exp\left[\ln\left(WT_s\right)-A^2/(2E_G) - \ln A\right] 
\]
The data rate (in bits/sec) is 
\[
R=\frac{\theta N}{T_c}\log_2\left(WT_s\right) = 
\frac{\theta N}{T_c}\log_2e\ln\left(WT_s\right)
\]
and the capacity \mbox{$\CAWGN = P / N_0 \log_2 e$}, so
\[
(\CAWGN / R) \ln (WT_s) = (PT_c) / (\theta N N_0) = \mathcal{E}/2N 
\]
Since \mbox{$A^2 = \alpha^2 \mathcal{E}/N$}, we obtain the bound
\begin{eqnarray*}
WT_sP(s_\mathrm{noise}\geq A) & \leq & \exp\left[\ln\left(WT_s\right) \right. \\
&&  - (\alpha^2 / E_G) (\CAWGN / R) \ln (WT_s) \\
&& \left.  - \ln \frac{A}{\sqrt{E_G}}\right] 
\end{eqnarray*}
Since \mbox{$E_G \rightarrow 1$}, the bound converges to 0 as long as \mbox{$ \alpha^2 > R / \CAWGN $}. This can be achieved for any data rate below the AWGN capacity. 

{\em Achieving Capacity:}
So far we have assumed $\alpha$ is a constant smaller than 1, so the communication system can achieve any rate below $\CAWGN$.
The duty cycle parameter $\theta$ must vary as $\frac{1}{\log W}$.
To achieve asymptotically $\CAWGN$, the parameter $\alpha$ must approach 1 as the bandwidth increases, and the following conditions need to be satisfied:
\begin{itemize}
\item \mbox{$(E_G - \alpha)^2 \log W \rightarrow \infty $} in probability (first error event)
\item \mbox{$ (\alpha^2 / E_G) (\CAWGN / R) \ge 1 $} with probability \mbox{$\rightarrow 1$} (third error event)
\end{itemize}
The exact choice of $\alpha$ depends on the rate at which $E_G$ converges to 1. 

{\em Summary of the Bound:}
The system uses IID symbols, a duty cycle $\theta$ and a threshold \mbox{$A=\alpha \sqrt{\mathcal{E}/N}=\alpha\sqrt{\frac{2PT_c}{N_0\theta N}}$} where \mbox{$\alpha \in (0,1)$}.

We calculated an upper bound on the error probability
\[
P(\mathrm{error}) \leq \mathrm{upper\ bound}(W,L,\frac{P}{N_0},\alpha, \theta)
\]
that converges to zero as \mbox{$W\rightarrow\infty$} if
\begin{itemize}
\item \mbox{$\frac{L^6}{W}\rightarrow0$} (second error event)
\item \mbox{$\theta\log W\sim\mathrm{const}$} (to ensure positive rate $R$)
\item \mbox{$\theta \log L\rightarrow 0$} (second error event)
\item \mbox{$\theta L\rightarrow 0$} (penalty for unknown gains)
\end{itemize}
If \mbox{$\frac{L}{\log W}\rightarrow0$} these the conditions can be realized simultaneously, namely it is possible to choose a duty cycle parameter $\theta$ that satisfies all the conditions.

Note that if the path delays are not known, the additional penalty~(\ref{eq:delaypenalty}) increases as $\theta L\log W$, which diverges, so the above proof is not useful.
\subsubsection{Estimation of Number of Overlap Terms}
\label{sec:numofpaths}

The number of possible path positions is \mbox{$L_m = \lfloor WT_d\rfloor$}. We assume that, 
over one coherence period, the $L$ delays are chosen uniformly at random among the $\binom{L_m}{L}$
possibilities. We prove that if the number of paths grows slowly enough, then with probability converging to 1
there will be at most one overlap between the set of delays and any of its translations.

\begin{Definition}
For any set $S \subset \bz$ and any integer $t \in \bz$, we denote by $S + t$ the translation of $S$ by $t$:
\mbox{$S + t = \{ s+t | s \in S\}$}.
\end{Definition}
$S$ corresponds to the received symbol when $x_1$ is transmitted, and $S+t$ corresponds to $x_{t+1}$.

For integers \mbox{$1 \le L \le L_m$} pick a random subset \mbox{$S \subset \{1,\ldots,L_m\}$} uniformly among all subsets of \mbox{$\{1,\ldots,L_m\}$} with $L$ elements.
Let $\pnl$ be the law of $S$; when there is no ambiguity we drop the subscripts and refer to it as $\bp$.

\begin{Theorem}\label{overlapmain}
Assume \mbox{$L^4 / L_m \rightarrow 0$} as \mbox{$L_m \rightarrow \infty$}, and a set $S$ is chosen according to $\pnl$. Then
\mbox{$ \pnl ( | S \cap (S+t) | > 1 {\mathrm{\ for \ some \ }}t \ne 0 ) \rightarrow 0. $}
\end{Theorem}

Note that $t$ can take both positive and negative values. We emphasize that the theorem says that with high probability, {\bf none} of the translates will have more than one overlap. The proof requires
the following

\begin{Lemma}\label{asubset}
Fix $t \ne 0$, and let $A$ be a set such that \mbox{$A \cup (A-t) \subset \{1,\ldots,L_m\}$}, and $|A|\le L$.
Then 
\begin{equation}
\bp(A \subset S \cap (S+t)) = [L]_a / [L_m]_a \le (L / L_m)^a
\label{eq:Pineq}
\end{equation}
where \mbox{$a = |A \cup (A-t)|$} and \mbox{$[x]_a = x(x-1) \ldots (x-a+1)$}.
\end{Lemma}
\begin{proof}
Clearly \mbox{$A \subset S \cap (S+t)$} is equivalent to \mbox{$A \subset S, (A-t) \subset S$}, hence $S$ has to contain $A \cup (A-t)$.
Hence $a$ elements of $S$ are fixed, while the remaining $L-a$ ones can be chosen in \mbox{$\binom{L_m-a}{L-a}$} ways.
The total number of subsets of \mbox{$\{1,\ldots,L_m\}$} with $L$ elements is \mbox{$\binom{L_m}{L}$}, hence
\[
\bp(A \subset S \cap (S+t)) = \binom{L_m-a}{L-a} / \binom{L_m}{L} = [L]_a / [L_m]_a
\]
The inequality~(\ref{eq:Pineq}) follows easily.
\end{proof}
\note If \mbox{$A \cup (A-t)$} is not a subset of \mbox{$\{1,\ldots,L_m\}$}, or if $|A|>L$, then the probability~(\ref{eq:Pineq}) is 0.\\


We are ready to obtain estimates. Fix $t>0$.
If \mbox{$|S \cap (S+t)| \ge 2$}, then \mbox{$S \cap (S+t)$} must contain \mbox{$A=\{i,j\}$} for some \mbox{$t+1 \le i<j\le L_m$}.
There are \mbox{$\binom{L_m-t}{2}$} such sets $A$.
Exactly $L_m-2t$ of them (namely $\{i,i+t\}$ for \mbox{$t+1\le i\le L_m-t$})
have \mbox{$|A \cup (A-t)| = 3$}; all the others have \mbox{$|A \cup (A-t)| = 4$}.
Hence
\begin{eqnarray*}
&& \bp(|S \cap (S+t)| \ge 2)  \\
&& \le  \sum_{t+1 \le i<j\le L_m} \bp(\{i,j\} \subset S \cap (S+t)) \\
&& \le (L_m - 2t) (L / L_m)^3 + \binom{L_m-t}{2} (L / L_m)^4 \\
&& \le \frac{1}{L_m^2}\left( L^3+L^4\right) 
\end{eqnarray*}
The same estimate holds for $t<0$. Hence
\begin{eqnarray*}
&& \bp ( | S \cap (S+t) | > 1 {\mathrm{\ for \ some \ }}t \ne 0 ) \\
&& \le \sum_{\stackrel{-L_m \le t \le L_m}{t \ne 0}} \bp ( | S \cap (S+t) | > 1) \\
&& \le  2L_m \frac{1}{L_m^2}\left( L^3+L^4\right) \\
&& \le  4 L^4 / L_m
\end{eqnarray*}
and the proof of Theorems~\ref{overlapmain} and~\ref{th:PPM_dknown} is complete.

%

\end{proof}

\subsection{When is the Channel Capacity Not Achieved?}
\label{sec:PPM_ub}

\setcounter{tempCounter}{\value{Theorem}}
\setcounter{Theorem}{2}
\addtocounter{Theorem}{-1}
\begin{Theorem_converse}
PPM systems with a lower bounded symbol time transmitting over a channel with Gaussian path gains that are unknown to the receiver, achieve \mbox{$C_\mathrm{PPM}\rightarrow0$} as \mbox{$W \rightarrow\infty$} if \mbox{$\frac{L}{\log W}\rightarrow \infty $}.
This result holds whether the receiver knows the path delays or it does not.
\end{Theorem_converse}
\setcounter{Theorem}{\value{tempCounter}}

The signals we consider are PPM with symbol time that may depend on the bandwidth, but cannot exceed the coherence period of the channel and cannot diminish (by assumption).
The symbol time is divided into positions separated by $\frac{1}{W}$.
Guard time may be used, no restriction is imposed over it, we use $T_\mathrm{symb}$ to denote the overall symbol time, that includes the guard time.
The signal transmitted over one coherence period is of the form:
\begin{eqnarray*}
X_i & = & \left\{ \begin{array}{ll} \sqrt{\frac{WT_\mathrm{symb}}{\theta}} & \mathrm{one\ position\ of\ each\ group\ of\ }\lfloor T_\mathrm{symb}W\rfloor\\ 
& \mathrm{with\ }n \scriptstyle{\lfloor T_\mathrm{symb} W\rfloor \leq i \leq n\lfloor T_\mathrm{symb} W\rfloor+\lfloor T_\mathrm{symb} W\rfloor-1} \\
& n=0,1,\dots, \left\lfloor \frac{T_c}{T_\mathrm{symb}}\right\rfloor-1 \ \ \ \ \ \mathrm{(symbol\ counter)}\\
\\
0 & \mathrm{other\ positions}
\end{array} \right. \\
&& i=0,1,\dots,\left\lfloor T_cW \right\rfloor-1 \ \ \ \ \ \mathrm{(position\ counter)} 
\end{eqnarray*}
The number of symbols transmitted over a single coherence period is \mbox{$N=\frac{T_c}{T_\mathrm{symb}}$}.
We assume that $N$ is a whole number, this assumption does not alter the result we prove here.
Duty cycle or any other form of infrequent transmission may be used over any time period. 
We analyze systems that use duty cycle over coherence periods, because this choice yields the highest data rate that serves as an upper bound.
The channel is composed of $L$ paths with independent and identically distributed Gaussian gains, and delays in the range $[0, T_d)$.

Edge effects between coherence periods are not considered, they may add a complication to the analysis, without contributing to the understanding of the problem or the solution. 

{\em Outline of the Proof of The Converse to Theorem~\ref{th:PPM_dknown}:}
The mutual information of the transmitted and received signals is upper bounded by the mutual information when the receiver knows the path delays.
This, in turn, is upper bounded in two ways:
the first is the PPM transmitted bit rate, and the second is based on the performance of a simple PPM system with no inter-symbol interference.
The proof is based on the conditions where the upper bound we calculate on the mutual information diminishes as the bandwidth increases.

\begin{proof}
We first point out that the mutual information of a system can only increase if the receiver is given information on the path delays:
\[
I(X;Y)\leq I(X;Y|D)
\]
We calculate an upper bound on PPM mutual information with a real Gaussian multipath channel, in [bits/sec]:
\begin{Proposition}
\label{prop:PPM_negative}
\begin{eqnarray}
I(X;Y|D)\ [\mathrm{b/s}]  & \leq  & \max_{0<\theta\leq 1} \min 
\left\{I_1(\theta), I_2(\theta)\right\} \label{eq:PPM_ub} \\
I_1(\theta)\ [\mathrm{b/s}] & \equiv & \frac{\theta\log_2 \left(WT_\mathrm{symb}\right)}{T_\mathrm{symb}} \nonumber \\
I_2(\theta)\ [\mathrm{b/s}] & \equiv & 
\frac{\theta W\left(T_d+T_\mathrm{symb}\right)}{2T_\mathrm{symb}} 
 \log_2\left( 1 + \frac{2PT_\mathrm{symb}}{\theta N_0W\left(T_d+T_\mathrm{symb}\right)}\right) \nonumber \\
&& -\frac{\theta L}{2 T_c} \log_2 \left(1+\frac{2PT_c}{\theta N_0L}\right) 
\end{eqnarray}
\end{Proposition}

{\em Discussion of Proposition~\ref{prop:PPM_negative}:}
The first part of the bound, $I_1(\theta)$, is an upper bound on the PPM bit rate for an uncoded system, it is a trivial upper bound on the mutual information.
$\theta$ is the fraction of time used for transmission, and the bound~(\ref{eq:PPM_ub}) is maximized over the choice of its value.
The second part, $I_2(\theta)$ depends on the number of channel paths $L$.

Using Proposition~\ref{prop:PPM_negative}, the converse to Theorem~\ref{th:PPM_dknown} follows simply:
The bound~(\ref{eq:PPM_ub}) is positive in the limit \mbox{$W\rightarrow\infty$} if both its parts are positive.
We note that the symbol time $T_\mathrm{symb}$ is lower bounded by a constant that does not depend on the bandwidth.
The first part, $I_1(\theta)$, is positive if the parameter $\theta$ is chosen so that \mbox{$\theta\log \left(WT_\mathrm{symb}\right)>0$}.
The second part $I_2(\theta)$ is positive in the limit of infinite bandwidth if $\theta L<\infty$.
If the environment is such that \mbox{$\frac{L}{\log W}\rightarrow\infty$}, the two conditions involving $\theta$ cannot be met simultaneously by any choice of fractional transmission parameter.
In this case, the bound~(\ref{eq:PPM_ub}) is zero in the limit of infinite bandwidth.

{\em Proof of Proposition~\ref{prop:PPM_negative}:} 
The first part of~(\ref{eq:PPM_ub}) follows simply from the fact that $I_1(\theta)$ is an upper bound on the transmitted data rate.
For any choice of fractional transmission parameter $\theta$:
\[
I(X;Y|D)\ [\mathrm{b/s}]\leq I_1(\theta)\ [\mathrm{b/s}]
\]

The second part of~(\ref{eq:PPM_ub}) is proven by comparing the mutual information of our system, with that of a hypothetical system that is easier to analyze.
The conditional mutual information $I(X;Y|D)$ is upper bounded using a hypothetical system that transmits the same symbols as the system we analyze, and receivers them without inter-symbol interference (ISI).
This is possible, for example, by using many pairs of transmitters and receivers, each pair transmitting symbols with long silence periods between them.
The transmitter--receiver pairs are located in such a way that each receiver can `hear' only its designated transmitter.
This hypothetical system operates over a channel identical to the one of the original system.
The difference between the original system and the hypothetical system is apparent in the number of different noise samples they face, the hypothetical system receives more noise, it processes $W\tilde{T}_c=WT_c\frac{T_\mathrm{symb}+T_d}{T_\mathrm{symb}}$ positions per coherence period.
In spite of this difference, the hypothetical system can achieve higher data rates and its mutual information is an upper bound on the mutual information in the original system.
We use $\tilde{X}$ and $\tilde{Y}$ to indicate the transmitted and received signals of this system.
\[
I(X;Y|D)\ [\mathrm{b/s}]\leq I(\tilde{X};\tilde{Y}|D)\ [\mathrm{b/s}]
\]

We now prove that for any choice of $\theta$
\[
I(\tilde{X};\tilde{Y}|D)\ [\mathrm{b/s}]\leq I_2(\theta)\ [\mathrm{b/s}]
\]

Each received symbol in the no-ISI system is composed of \mbox{$W\left(T_\mathrm{symb}+T_d\right)$} chips, $L$ of them corresponding to the channel paths.
All output positions have additive Gaussian noise of variance one.
The mutual information is given by
\begin{equation}
I(\tilde{X};\tilde{Y}|D)\ [\mathrm{b/s}]=\frac{\theta}{T_c}\left[H(\tilde{Y}|D)-H(\tilde{Y}|\tilde{X},D)\right] \label{eq:I3}
\end{equation}

We start with the first part of~(\ref{eq:I3}):
\[
H(\tilde{Y}|D)\leq \sum_{i=1}^{W\tilde{T}_c}H(\tilde{Y}_i|D)\leq \sum_{i=1}^{W\tilde{T}_c}H(\tilde{Y}_i)
\]
\begin{eqnarray*}
\tilde{Y}_i & \sim & \left\{ \begin{array}{ll}
N\left(0, 1+\frac{2PT_\mathrm{symb}}{\theta N_0L}\right) & \mathrm{prob}\ p_i\\
N(0, 1) & \mathrm{prob}\ 1-p_i
\end{array}
\right. \\
&& i=1, \dots, W(T_d+T_\mathrm{symb})
\end{eqnarray*}
$p_i$, the probability of receiving signal energy in the \thd{$i$} position, depends on the distribution of transmitted symbols, but there are exactly $L$ positions in the received symbol the contain a path, thus
\[
\sum_{i=1}^{W\left(T_d+T_\mathrm{symb}\right)}p_i=L
\]
and each probability value satisfies $0\leq p_i\leq 1$.
\begin{eqnarray*}
\sigma_{\tilde{Y}_i}^2 
& = & E\left[\tilde{Y}_i^2\right]  =  1-p_i+
p_i
\left(1+\frac{2PT_\mathrm{symb}}{\theta N_0L}\right) \\
& = &   1+
p_i\frac{2PT_\mathrm{symb}}{\theta N_0L} 
\end{eqnarray*}
\begin{eqnarray*}
H(\tilde{Y}_i) & \leq & \frac{1}{2}\log(2\pi e \sigma_{\tilde{Y}_i}^2) \\
& = & \frac{1}{2}\log\left(2\pi e \left( 1 + p_i\frac{2PT_\mathrm{symb}}{\theta N_0L}  \right)\right)
\end{eqnarray*}
\begin{eqnarray*}
H(\tilde{Y}|D) & \leq & \sum_{i=1}^{W\tilde{T}_c}H(\tilde{Y}_i) \\
& \leq & \sum_{T_c/T_\mathrm{symb} \mathrm{\ symbols}}\sum_{i=1}^{W(T_d+T_\mathrm{symb})} \frac{1}{2}\log\left(2\pi e \left( 1 + p_i\frac{2PT_\mathrm{symb}}{\theta N_0L}  \right)\right) \\
& = & \frac{W\tilde{T}_c}{2}\log\left(2\pi e \right)+\sum_{\mathrm{symbols}}\sum_{i=1}^{W(T_d+T_\mathrm{symb})} \frac{1}{2}\log\left( 1 + p_i\frac{2PT_\mathrm{symb}}{\theta N_0L}  \right) \\
& = & \frac{W\tilde{T}_c}{2}\log\left(2\pi e \right)+\frac{T_c}{2T_\mathrm{symb}}\sum_{i=1}^{W\left(T_d+T_\mathrm{symb}\right)} \log\left( 1 + p_i\frac{2PT_\mathrm{symb}}{\theta N_0L}\right) 
\end{eqnarray*}
Using the concavity of the $\log$ we get
\begin{eqnarray*}
H(\tilde{Y}|D) & \leq & 
\frac{W\tilde{T}_c}{2}\log\left(2\pi e \right)+\frac{WT_c\left(T_d+T_\mathrm{symb}\right)}{2T_\mathrm{symb}} \log\left( 1 + \frac{2PT_\mathrm{symb}}{\theta N_0W\left(T_d+T_\mathrm{symb}\right)}\right) 
\end{eqnarray*}

Now for the second part of~(\ref{eq:I3}).
For $N$ transmitted symbols, the $W\tilde{T}_c$ received values are distributed as follows, when the input $\tilde{X}$ and the delays $D$ are known:
\begin{itemize}
\item  $W\tilde{T}_c-NL$ positions are IID Gaussians $\sim N(0,1)$. The receiver knows which positions contain only noise, and which have signal as well.
\item 
The number of positions with some signal is $NL$.
These values are divided into groups of size $N$, each corresponding to a single path.
Each group (at known positions) is independent of the other groups and its distribution is \mbox{$\sim N(0, \Lambda)$} where
\begin{eqnarray*}
\Lambda & = & \frac{2PT_\mathrm{symb}}{\theta N_0L}+\left(\begin{array}{ccc} 1  & & 0 \\ & \ddots & \\0 & & 1 \end{array}\right) 
\end{eqnarray*}
\end{itemize}
\begin{eqnarray}
-H(\tilde{Y}|\tilde{X},D)& = & -\frac{W\tilde{T}_c-NL}{2}\log(2\pi e) \nonumber \\
&& -\frac{L}{2}\log\left((2\pi e)^N\left|\Lambda\right|\right) \label{eq:I7}
\end{eqnarray}

The determinant $|\Lambda|$ is the product of the eigenvalues of
$\Lambda$: $1$ with multiplicity $N-1$ and
\mbox{$\left(1+\frac{2PT_\mathrm{symb}N}{\theta N_0L}\right)$} with multiplicity one
\[
|\Lambda|=1+\frac{2PT_\mathrm{symb}N}{\theta N_0L}
=1+\frac{2PT_c}{\theta N_0L}
\]
Simple manipulation yields
\begin{eqnarray*}
-H(\tilde{Y}|\tilde{X},D) 
& = & -\frac{W\tilde{T}_c}{2}\log(2\pi e) \\
&& -\frac{L}{2} \log \left(1+\frac{2PT_c}{\theta N_0L}\right)
\end{eqnarray*}
We note that this expression depends on $\mathrm{SNR}_\mathrm{est}$; 
It is similar to expressions in~\cite{telatar_2000}, Section III.C.

Combining the two parts into~(\ref{eq:I3}) we get
\begin{eqnarray*}
I(X;Y|D)\ [\mathrm{b/s}]   & \leq  & I(\tilde{X};\tilde{Y}|D)\ [\mathrm{b/s}]   \\
& = &  \left[H(\tilde{Y}|D)-H(\tilde{Y}|\tilde{X},D)\right]\frac{\theta}{T_c} \\
& \leq &  \left[
\frac{W\tilde{T}_c}{2}\log_2\left(2\pi e \right)+\frac{WT_c\left(T_d+T_\mathrm{symb}\right)}{2T_\mathrm{symb}} \log_2\left( 1 + \frac{2PT_\mathrm{symb}}{\theta N_0W\left(T_d+T_\mathrm{symb}\right)}\right)  \right. \\
&& \left. -\frac{W\tilde{T}_c}{2}\log_2(2\pi e)-\frac{L}{2} \log_2 \left(1+\frac{2PT_c}{\theta N_0L}\right)
\right] 
\frac{\theta}{T_c}\\
& = &  
\frac{\theta W\left(T_d+T_\mathrm{symb}\right)}{2T_\mathrm{symb}} 
 \log_2\left( 1 + \frac{2PT_\mathrm{symb}}{\theta N_0W\left(T_d+T_\mathrm{symb}\right)}\right) \\
&& -\frac{\theta L}{2 T_c} \log_2 \left(1+\frac{2PT_c}{\theta N_0L}\right) \\
& = &  I_2(\theta)
\end{eqnarray*}
Note that in the infinite bandwidth limit, if $\theta L\rightarrow\infty$ than also $\theta W\rightarrow\infty$ and $I_2(\theta)$ converges to zero:
\[
I_2(\theta)\longrightarrow
\frac{\theta}{2T_\mathrm{symb}}\sum_{i=1}^{W\left(T_d+T_\mathrm{symb}\right)} p_i\frac{2PT_\mathrm{symb}}{\theta N_0L}\log_2 e  
 -\frac{\theta L}{2 T_c} \frac{2PT_c}{\theta N_0L} \log_2e=0
\]

\end{proof}

\section{Conclusion}
\label{sec:conclusion}

This paper revealed the importance of the rate of increase of the apparent number of paths as the bandwidth increases.
This rate of increase determines the behavior of channel uncertainty as bandwidth increases, thus determining the spectral efficiency required of a communication system, in order to achieve channel capacity in the limit of large bandwidth.
Practical mobile communication systems handle channel uncertainty by occasional estimation of the channel, usually assisted by pilot (training) signals.
The rate of repeating the pilot signal is determined by the rate of channel variation and the required accuracy of estimation.
This paper showed that systems with a low spectral efficiency, and in particular PPM systems, have to sacrifice a significant portion of their data-rate in order to obtain channel estimation.
When operating over difficult channels, namely channels where the number of paths increases quickly with the bandwidth, a PPM system has to devote an increasing fraction of its resources to channel estimation, until in the limit of infinite bandwidth it cannot communicate.

The key to understanding the wideband operation over a radio channel is the behavior of channel uncertainty as bandwidth increases.
The channel model we used is a simplistic one, where the channel paths are similar so each represents the same amount of channel uncertainty.
Realistic channel models are more complex, so it is harder to characterize their channel uncertainty and its dependence on system bandwidth.
Wideband radio channel models show a spread of the impulse response energy over a number of paths, that may not be of equal amplitude. 
The dependence of channel uncertainty on the bandwidth depends on the rate of increase of the number of significant channel paths as the bandwidth increases, and we expect similar results to ours, where the number of significant paths replaces the number of paths $L$ we used.

\appendix

This section shows that for IID Gaussians $\left\{r_i\right\}_{i=-\infty}^{\infty}\sim N(0,\frac{1}{\theta})$, with the empirical correlation defined by:
\[
C(m,n)=\frac{\theta}{K_c}\sum_{i=0}^{K_c-1} r_{i-m}r_{i-n}
\]
We have
\begin{equation}
\sum_{m=1}^L \sum_{n=1}^L E_x \left| C(m,n)-\delta_{mn}\right| \leq  \frac{2L}{\pi \sqrt{K_c}} +\frac{L^2-L}{K_c}
\label{eq:ssE}
\end{equation}

We first look at $C(m,n)$ for $m\neq n$, and use the following inequality, that holds for any random variable:
\[
E\left| c \right|\leq \sqrt{E\left[ c^2\right]}
\]
in our case:
\begin{eqnarray}
E\left| C(m,n) \right| & \leq & \sqrt{E\left[ C(m,n)^2\right]} \nonumber \\
& = & \left\{\frac{\theta^2}{K_c^2}E \left[\left(\sum_{i=0}^{K_c-1} r_{i-m}r_{i-n}\right) \right. \right. \nonumber \\
&&\left.\left. \left(\sum_{j=0}^{K_c-1} r_{j-m}r_{j-n}\right) \right]\right\}^{1/2}\nonumber \\
& = & \frac{\theta}{K_c} \sqrt{E\left[ \sum_{i=0}^{K_c} r_{i-m}^2r_{i-n}^2\right]}^{1/2} \nonumber \\
& = & \frac{1}{K_c} \label{eq:p1}
\end{eqnarray}

Now for the case $m=n$
\begin{eqnarray*}
E\left[r_{i-m}r_{i-n}\right] & = & E\left[r_{i-m}^2\right] = \frac{1}{\theta} \\
E\left[r_{i-m}^4\right] & = & \frac{3}{\theta^2}  
\end{eqnarray*}
The fourth moment is so because $r_i$ is Gaussian.

Using the central limit theorem (holds as $r_i$ are independent)
\[
\sum_{i=0}^{K_c-1} \left(r_{i-m}^2-\frac{1}{\theta}\right)\sim N\left(0,\frac{2K_c}{\theta^2} \right)
\]
\begin{equation}
E\left|C(m,m)-1\right|=\frac{\theta}{K_c} E\left|\sum_{i=0}^{K_c-1} \left(r_{i-m}^2-\frac{1}{\theta}\right)\right|= \frac{2}{\sqrt{\pi K_c}}
\label{eq:p2}
\end{equation}

Taking~(\ref{eq:p1}) and~(\ref{eq:p2}) into~(\ref{eq:ssE}) gives:
\begin{eqnarray*}
&& \sum_{m=1}^L\sum_{n=1}^L E\left|C(m,n)-\delta_{mn}\right|   \\
&&  =LE\left|C(m,m)-1\right|+\left(L^2-L\right)E\left|C(m,n)\right|_{m\neq n} \\
&& \leq \frac{2L}{\sqrt{\pi K_c}} +\frac{L^2-L}{K_c}
\end{eqnarray*}

\bibliography{IEEEabrv,refs}

\begin{thebibliography}{10}
\providecommand{\url}[1]{#1}
\csname url@rmstyle\endcsname
\providecommand{\newblock}{\relax}
\providecommand{\bibinfo}[2]{#2}
\providecommand\BIBentrySTDinterwordspacing{\spaceskip=0pt\relax}
\providecommand\BIBentryALTinterwordstretchfactor{4}
\providecommand\BIBentryALTinterwordspacing{\spaceskip=\fontdimen2\font plus
\BIBentryALTinterwordstretchfactor\fontdimen3\font minus
  \fontdimen4\font\relax}
\providecommand\BIBforeignlanguage[2]{{%
\expandafter\ifx\csname l@#1\endcsname\relax
\typeout{** WARNING: IEEEtran.bst: No hyphenation pattern has been}%
\typeout{** loaded for the language `#1'. Using the pattern for}%
\typeout{** the default language instead.}%
\else
\language=\csname l@#1\endcsname
\fi
#2}}

\bibitem{kennedy_book}
R.~S. Kennedy, \emph{Fading Dispersive Communication Channels}.\hskip 1em plus
  0.5em minus 0.4em\relax Wiley \& Sons, 1969.

\bibitem{gallager_book_sec8.6}
R.~G. Gallager, \emph{Information Theory and Reliable Communication}.\hskip 1em
  plus 0.5em minus 0.4em\relax Wiley \& Sons, 1968, section 8.6.

\bibitem{telatar_2000}
I.~E. Telatar and D.~N.~C. Tse, ``Capacity and mutual information of wideband
  multipath fadng channels,'' \emph{IEEE Transactions on Information Theory},
  vol.~46, no.~4, pp. 1384--1400, July 2000.

\bibitem{medard_2002}
M.~M\'{e}dard and R.~G. Gallager, ``Bandwidth scaling for fading multipath
  channels,'' \emph{IEEE Transactions on Information Theory}, vol. 2002, no.~4,
  pp. 840--852, April 2002.

\bibitem{subramanian_2002}
V.~G. Subramanian and B.~Hajek, ``Broad-band fading channels: Signal burstiness
  and capacity,'' \emph{IEEE Transactions on Information Theory}, vol.~48,
  no.~4, pp. 809--827, April 2002.

\bibitem{zheng_OnChannel_2003}
L.~Zheng, M.~M. adn David~Tse, and C.~Luo, ``On channel coherence in the low
  {SNR} regime,'' in \emph{41st Allerton Conference on Communication, Control
  and Computing}, vol.~1, October 2003, pp. 420--429.

\bibitem{rusch_2002}
L.~Rusch, C.~Prettie, D.~Cheung, Q.~Li, and M.~Ho, ``Characterization of {UWB}
  propagation from 2 to 8 {GHz} in a residential environment,'' \emph{IEEE
  Journal on Selected Areas in Communications}.

\bibitem{pendergrass_2002}
M.~Pendergrass, ``Empirically based statistical ultra-wideband model,''
  \emph{IEEE 802.15 Working Group for Wireless Area Networks (WPANs)}, July
  2002, contribution 02/240, available at
  http://grouper.ieee.org/groups/802/15/pub/2002/Jul02/02240r1p802-15\_SG3a-Em%
pirically\_Based\_UWB\_Channel\_Model.ppt.

\bibitem{verdu_2002}
S.~Verd\'{u}, ``Spectral efficiency in the wideband regime,'' \emph{IEEE
  Transactions on Information Theory}, vol.~48, no.~6, pp. 1319--1343, June
  2002.

\bibitem{zheng_Channel_2004}
L.~Zheng, D.~Tse, and M.~M\'{e}dard, ``Channel coherence in the low snr
  regime,'' in \emph{International Symposium on Information Theory
  (ISIT)}.\hskip 1em plus 0.5em minus 0.4em\relax IEEE, June-July 2004, p. 413.

\bibitem{zheng_OnThe_2004}
------, ``On the costs of channel state information,'' in \emph{Information
  Theory Workshop (ITW)}.\hskip 1em plus 0.5em minus 0.4em\relax IEEE, October
  2004, pp. 423--427.

\end{thebibliography}
\end{document}